\documentclass[twocolumn,twocolappendix,tighten]{aastex63}
\usepackage{amsmath}
\newcommand{\HI}{H\textsc{i}\,\,}
\newcommand{\HII}{H\textsc{i}} %same as above but without space in the end
\newcommand{\dHI}{\textup{HI}\,}

\newcommand{\beq}{\begin{equation}}
\newcommand{\eeq}{\end{equation}}
\def\Mpc{\, h^{-1} \, {\rm Mpc}}
\def\kMpc{\, h \, {\rm Mpc}^{-1}}
\def\Kpc{\, h^{-1} \, {\rm kpc}}

\def\x{{\bf x}}
\graphicspath{{./}{figures/}}
\begin{document}

\title[HInet: Neutral Hydrogen with neural networks]{HInet: Generating neutral hydrogen from dark matter with neural networks}
\correspondingauthor{Digvijay Wadekar}
\email{jay.wadekar@nyu.edu}

\author[0000-0002-2544-7533]{Digvijay Wadekar}
\affiliation{Center for Cosmology and Particle Physics, Department of Physics,
New York University, New York, NY 10003, USA}

\author[0000-0002-4816-0455]{Francisco Villaescusa-Navarro}
\affiliation{Department of Astrophysical Sciences, Princeton University, Peyton Hall, Princeton NJ 08544-0010, USA}
\affiliation{Center for Computational Astrophysics, Flatiron Institute, 162 5th Avenue, 10010, New York, NY, USA}

\author[0000-0002-1068-160X]{Shirley Ho}
\affiliation{Department of Astrophysical Sciences, Princeton University, Peyton Hall, Princeton NJ 08544-0010, USA}
\affiliation{Center for Computational Astrophysics, Flatiron Institute, 162 5th Avenue, 10010, New York, NY, USA}
\affiliation{Department of Physics, Carnegie Mellon University, Pittsburgh, PA 15217, USA}

\author[0000-0003-3544-3939]{Laurence Perreault-Levasseur}
\affiliation{Center for Computational Astrophysics, Flatiron Institute, 162 5th Avenue, 10010, New York, NY, USA}
\affiliation{ Department of Physics, Univesit\'e de Montr\'eal, Montr\'eal, Canada}
\affiliation{Mila - Quebec Artificial Intelligence Institute,
Montr\'eal, Canada}

\begin{abstract}

Upcoming 21cm surveys will map the spatial distribution of cosmic neutral hydrogen (HI) over very large cosmological volumes. In order to maximize the scientific return of these surveys, accurate theoretical predictions are needed. Hydrodynamic simulations currently are the most accurate tool to provide those predictions in the mildly to non-linear regime. Unfortunately, their computational cost is very high: tens of millions of CPU hours. We use convolutional neural networks to find the mapping between the spatial distribution of matter from N-body simulations and HI from the state-of-the-art hydrodynamic simulation IllustrisTNG. Our model performs better than the widely used theoretical model: Halo Occupation Distribution (HOD) for all statistical properties up to the non-linear scales $k\lesssim1$ h/Mpc. Our method allows the generation of 21cm mocks over very big cosmological volumes with similar properties as hydrodynamic simulations.
\end{abstract}

\keywords{large-scale structure, neutral hydrogen, deep learning}

\section{Introduction}
 Different astronomical surveys have allowed us to quantify the amount and properties of several fundamental quantities like the age, geometry and expansion rate of the Universe, and the amount of dark matter and dark energy. Some of the largest surveys in the past have been spectroscopic surveys of galaxies, which have mapped the Universe at low redshifts.
In future surveys, we want to observe the Universe at high-$z$ because the cosmic volume is larger and the theoretical predictions at high-$z$ are relatively easier as the density field is more linear.

The traditional technique of getting spectra of individual galaxies becomes harder to apply at high-$z$ as the galaxies become fainter and sparser. %\Paco{This is nice, but all these sentences seem a bit disconnected. You need to find a way to connect them. My suggestion, write in one sentence what you want to do/what is the problem, and then expand from there. I think you can focus this as galaxy redshift surveys (or astronomical surveys in general) as a way to constrain cosmological parameters and learn about fundamental physics.}
One of the most promising alternative techniques to observe the high-$z$ Universe is Intensity Mapping (IM)  \citep{BhaNatSet0103, BhaSet0112,ChaPen08,Pet09,Pul14} %\shir{maybe some references to Anthony Pullen, Ue-Li Pen, Jeff Peterson,Olivier Dore, tzu-Ching Chang's work?}.
The advantage of IM over the traditional techniques is that it does not rely on resolving point sources but instead measures the emission from many unresolved galaxies tracing the cosmic web in redshift space.

In this paper we focus our attention on IM of the 21cm line from cosmic neutral hydrogen (hereafter \HI).
It is worth noting that 21cm IM is not just restricted to the high-$z$ but is applicable over a wide range of redshifts
($z=0$ to $z\simeq20$). 21cm surveys represent a different way to observe the Universe, and they enable new cross-correlations with surveys at other wavelengths, which are very effective to mitigate systematic effects. Besides traditional bounds on cosmological parameters \citep{Bul15,VilAloVie17}, 21cm surveys can be used to improve our knowledge on the sum of neutrino masses \citep{VilBulVie15}, warm DM \citep{CarVilVie15}, modified gravity \citep{CarCorVie17}, primordial non-Gaussianity \citep{KarSloLig19} and axion DM \citep{BauMarHlo20}, among many other things.

In this work, we focus our attention on the the post-reionization regime ($z<6$) of 21 cm IM\footnote{A part of this regime (e.g. $z\in[4-6]$) has never been mapped with surveys before.}, for which various radio surveys are planned or are already collecting data:
Canadian Hydrogen Intensity Mapping Experiment (CHIME)
\footnote{\url{http://chime.phas.ubc.ca/}}
Giant Meterwave Radio Telescope (GMRT) \footnote{\url{http://gmrt.ncra.tifr.res.in/}},
HIRAX (The Hydrogen Intensity and Real-time Analysis eXperiment)
\footnote{\url{https://www.acru.ukzn.ac.za/~hirax/}}, TIANLAI \footnote{\url{http://tianlai.bao.ac.cn}}, Five-hundred-meter Aperture Spherical Telescope (FAST) \footnote{\url{https://fast.bao.ac.cn/en/}},
ASKAP \footnote{\url{http://www.atnf.csiro.au/projects/askap/index.html}}, MeerKAT \footnote{\url{http://www.ska.ac.za/meerkat/}}, PUMA \footnote{\url{https://www.puma.bnl.gov/}},
SKA (The Square Kilometer Array)
\footnote{\url{https://www.skatelescope.org/}}. 

One of the major aims of these surveys is to accurately constrain the value of the cosmological parameters. In order to achieve this, accurate theoretical predictions are needed to extract the cosmological information from the collected data. In the linear regime, these predictions are easy to obtain from analytical models and are accurate. However, there is a large amount of cosmological information that lies beyond the linear scales, particularly at low-$z$. In this regime, one avenue to obtain such accurate predictions is from hydrodynamic simulations.  
% \shir{how much non-linear or semi-nonlinear info do we still have beyond redshift of 4?}

Current state-of-the-art hydrodynamic simulations have a very high computational cost and they simulate a limited cosmological volume. For example, simulating the $(75 \Mpc)^3$ IllustrisTNG box required 18 million CPU hours \citep{NelPilAnn19,PilSprNel1801,WeiSprHer1703}. In order to make robust predictions for upcoming astronomical surveys we need to simulate much larger cosmological volumes, of the order of tens to hundreds of $(\textup{Gpc/h})^3$ \citep{ModCasFen1909}. Such large mock simulations would help us in various ways: 1) to study the effects of various observational systematics on the statistical properties of the tracers, 2) to obtain theoretical predictions for different cosmologies, 3) to determine which summary statistics are the most appropriate to constrain the value of different cosmological parameters,
4) to quantify the cosmic variance in the surveys (i.e, to compute the covariance matrix). 
One way to simulate large \HI volumes is to first generate dark matter (DM) fields using the relatively less expensive DM-only simulations. We then need quick and reliable methods to `paint' \HI directly on the DM field. We now discuss some promising techniques in this regard.

\subsection{Traditional emulation techniques}
% \LL{Maybe this section shouldnt be called HOD since it also discusses SAM, SHAM and EFTofLSS}
One the most popular theoretical techniques used to make mock baryonic simulations is called the Halo Occupation Distribution (HOD). HOD was first used to probabilistically model the number of galaxies residing in a host halo \citep{ScoSheHui01,Sel00,PeaSmi00,BerWei02}.
The HOD technique assumes that the properties of various baryonic structures inside a halo are governed \emph{solely by the halo mass}, and ignores all other halo properties. The HOD technique therefore assumes a simple parametric relation between the halo mass and the baryonic properties and uses hydrodynamical simulations (and observations, if available) to calibrate the parameters in this relation. Recent applications of the HOD technique to \HI have been in initial field reconstruction and testing the UV background effect on \HI maps \citep{ModCasFen1909,ModWhiSlo1911} based on the HOD model of \defcitealias{VilGenCas1810}{VN18} Villaescusa-Navarro et al.\ (\citeyear{VilGenCas1810}, hereafter \citetalias{VilGenCas1810}). 

However, HOD ignores all environmental effects on the \HI abundance and clustering.
It is important to note that numerical simulations have shown that cosmological properties like the clustering of halos and galaxies are affected by properties other than halo mass like halo environment, halo concentration, spin and velocity anisotropy and others \citep{Wec06, Dal08,ParHahShe18,HadBosEis20}. This phenomenon referred to as \emph{assembly bias} or secondary bias \citep{SheTor04,Gao05}. %\Paco{\cite{VilGenCas1810} showed that halo clustering is also affected by the halo HI mass, i.e. by internal galaxy properties.} \Jay{Didn't want to add this here as we talk about it after Eq.(1)}

Other techniques can be used to make mock baryonic simulations, such as
Subhalo Abundance Matching (SHAM) and semi-analytic models (SAM). SHAM involves assigning the highest \HI mass to the most massive halos and vice versa but it relies on multiple assumptions like more massive baryonic structures are hosted by the most massive halos and a monotonic relation ,which is free of scatter, exists between masses of baryonic structures and halo masses.
SAM, on the other hand, uses a set of simplified equations to model the key baryonic processes in the hydrodynamic simulations (for example, see \cite{Ben12}).

It is worth mentioning that, apart from using simulations, there are also proposed perturbative forward model techniques to evolve the tracer fields directly from linear initial conditions by using biasing schemes \citep{SchSimZal19, ModCheWhi20}
(or modeling tracer fields by combining biasing schemes and $N$-body simulations \citep{Sin20}).

\subsection{Convolutional neural networks}
Convolutional neural networks (CNNs) have been recently applied to numerous areas of physical research \citep{CarCirCra19} and have a lot of potential applications to cosmology
\citep{Nta19,RavOliFro1711,ZamOkaVil1904,ZhaWanZha1902,HeLiFen1907,GuiReyVil1910, ModFenSel18,Kod20}. An important property of CNNs which makes them useful for cosmological applications is that they are translationally equivariant.
In this paper, we use a modified version of a neural network architecture called U-Net to generate mock 3D \HI fields from a given DM field. Although we focus on \HI in the post-reionization regime in this paper, it is worth mentioning some of the recent work on using neural networks to study the \HI field during the epoch of reionization \citep{Shi17,Gil19,Cha19,Lis20,Kwo20,Man20,Has20, VilVil20}.

Let us now briefly highlight some of the advantages of neural networks over the above traditional approaches.
The HOD formalism typically assumes that the spatial \HI density $\rho_{\rm HI}(\x)$ at a particular point inside a halo 
only depends on the mass of the halo and the distance to its center:
%\footnote{$\rho_{\rm HI}(\x)$ for HOD also depends on the distance to the center of the halo if one includes the one-halo term.}:

\beq
\rho_\textup{HI}(\x)= f(M_{\rm halo},|\x-\x_{\rm center}|)
\eeq

% on the local DM density\footnote{\Paco{This is simplified version. In reality, the HI density at some point will depends on whether a dark matter halo exists at that location, and then on its mass and distance to the halo center.}} $\rho_{\rm m}(\x)$ at that location, i.e 
% \beq
% \rho_\textup{HI}(\x)=f(\rho_{\rm m}(\x))
% \eeq
However, \citetalias{VilGenCas1810} showed that including only the halo mass is not enough for precisely modeling the clustering of the \HI field.
A more general model for \HI field should also include the information on the environment of the halo and can be roughly intuited as 
%and therefore a more appropriate relation between the spatial distribution of HI and the matter should be:
\begin{equation}
    \rho_{\rm HI}(\x)=g(\rho_{\rm m}(\x), \rho_{\rm m}(\x'))
\label{eq:f2}\end{equation}
where $\rho_{\rm m}(\x')$ is the matter density at points in the neighborhood\footnote{We will precisely define the extent of the neighborhood later in Section~\ref{sec:training}.} of $\x$.
% \shir{wonder if we should explicitly say that x' contains all points in the X Mpc/h sphere of x?}
% \shir{what is the exact limit we have here? } \Jay{resolved in footnote as didn't want to get bogged down in details at this point.}
Neural networks are universal fitting functions \citep{HorSti89} and can be used to accurately approximate the function $g$ in Equation~\ref{eq:f2}; this is the goal of our paper.

The paper is organized as follows. In Section~\ref{sec:data}, we briefly describe the hydrodynamical simulations that we have used. We then present the specific architecture of our machine learning model and the method used in Section~\ref{sec:method}. We discuss the parameters of our benchmark HOD model in Section~\ref{sec:HOD}. We present our results in Section~\ref{sec:results}. Finally, we discuss our results in Section~\ref{sec:discussion} and conclude in Section \ref{sec:conclusions}.
%In Appendix \ref{apx:details}, we present supplementary details about training the neural network. 

\section{Data}
\label{sec:data}
The data we use to train, validate and test our network is obtained from the TNG100-1 simulation produced by the IllustrisTNG collaboration \citep{PilSprNel1801}\footnote{\url{https://www.tng-project.org/data/}}. That simulation is one of the current state-of-the-art hydrodynamical simulations and includes a wide range of relevant physical effects, such as radiative cooling, star formation, metal enrichment, supernova and AGN feedback, and magnetic fields. In this work, we choose to model the \HI field at low redshift: $z=1$; this is because modeling any baryonic field is more challenging at lower redshifts due the density fluctuations being relatively non-linear. We therefore expect our neural network method to perform even better at higher redshifts.

The side length of the simulated box is 75 h$^{-1}$ Mpc. We first compute the \HI density field by assigning \HI masses of gas cells to a grid of $2048^3$ cells using the cloud-in-cell (CIC) interpolation scheme. The spatial resolution of the DM and \HI fields is therefore very high $\sim 35 \Kpc$.
TNG provides both the hydrodynamical simulation output as well as the computationally cheaper dark matter only simulation (TNG100-1-DM), evolved from the same initial conditions. Our goal is to train a neural network such that it can produce the \HI field from the DM only simulation. The network performs the mapping in 3D, at a fixed redshift.

\subsection{Data preprocessing}
\label{sec:DataProcessing}
The overdensity in the \HI field, $\delta_\dHI=\rho_{\rm HI}/\bar{\rho}_{\rm HI}-1$, varies in the TNG100-1 simulation across $\sim$9 orders of magnitude. Because the resolution of the TNG100-1 simulation is much higher than the one expected from upcoming surveys, we smooth the \HI data with a Top-Hat filter with a smoothing radius of 300 $h^{-1}{\rm kpc}$. This has a two-fold advantage: First, the grid resolution for the \HI field is lowered to 140 $\Kpc$, which reduces the size of the dataset. Second, the dynamical range over which the \HI density field varies is reduced: $\delta_\dHI$ varies over three orders of magnitude and this reduces the sparsity problem which we later discuss in section~\ref{sec:method}. Note that we did not change the resolution of the input DM field, in order to use as much information in the DM field as possible. Because the training of deep learning models is facilitated when the input data is in the $\mathcal{O}(1)$ range, we further perform the scaling:
\beq\begin{split}
\tilde{\delta}_\dHI(\x)&\equiv\frac{1}{2}(1+\delta_\dHI(\x))^{0.2} \\
\tilde{\delta}_\textup{DM}(\x)&\equiv\frac{1}{5}(1+\delta_\textup{DM}(\x))^{0.1}\, ,
\end{split}\label{eq:DataTransform}\eeq
where $\delta_\dHI$ is the smoothed \HI field. The above rescaling get both $\tilde{\delta}_\dHI$ and $\tilde{\delta}_\textup{DM}$ to be in the nearly in the range [0, 3]. We used a power law instead of a logarithm because the power law distribution has a flatter tail for high values of $\delta_\dHI$. We discuss why having a flatter tail is important in Section~\ref{sec:DataSparsity}.
%---------------------------------------------------------------------------------------------
\section{Methods}
\label{sec:method}

\subsection{Choice of network architecture}
The deep neural network architecture used in this paper is inspired by the Deep Density Displacement Model (D$^3$M) of \cite{HeLiFen1907}.  D$^3$M is the generalization of the standard U-Net which was first proposed by \cite{Unet} for use in medical applications. Convolutional neural networks, like the U-Net, naturally provide properties which are relevant for our problem such as translational invariance. Variations of the D$^3$M model have been used for large scale structure applications like learning galaxy modeling and neutrino effects in cosmology \citep{YipZha19,ZhaWanZha1902,GuiReyVil1910}.
The network architecture we use in this work is shown in Figure~\ref{fig:Architecture}, and further details are presented in Appendix~\ref{apx:details}.

\begin{figure*}
    \centering
        \includegraphics[scale=0.45,keepaspectratio=true]{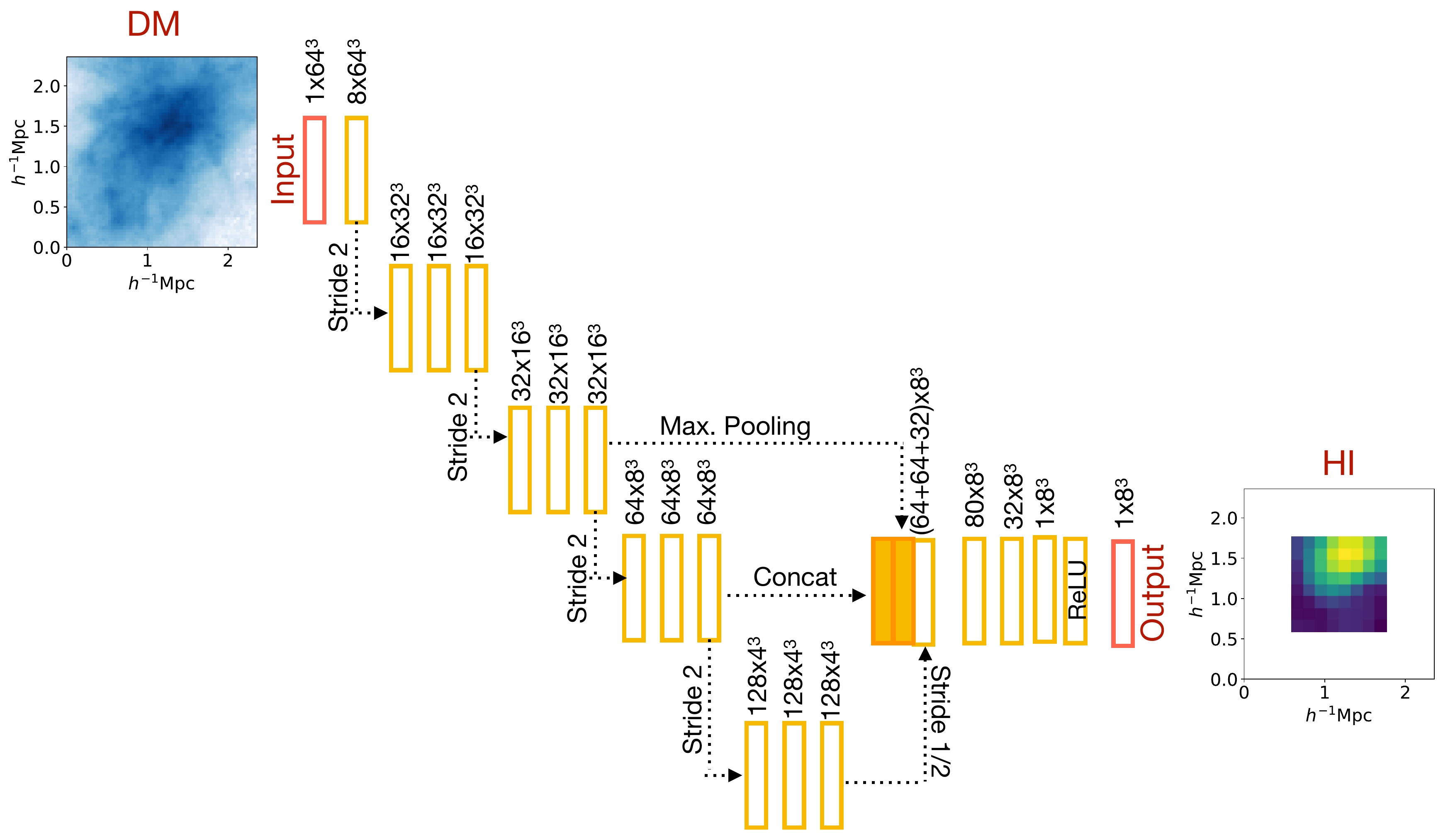}
\caption{This scheme shows the architecture we use to find the mapping from dark matter to HI in 3D. Labels represent [number of channels $\times$ (number of voxels)$^3$]. Further details can be found in Appendix \ref{apx:details}. Notice that the volume of the \HI output is $1/8^\textup{th}$ of the input DM volume, which enables inclusion of some of the non-local (environmental) information to predict the \HI field.}
    \label{fig:Architecture}
\end{figure*}

\subsection{Challenge of data sparsity}
\label{sec:DataSparsity}
Let us discuss an important challenge we face when working with HI data from simulations. The most relevant summary statistics for 21cm IM, e.g. the HI power spectrum or the HI PDF, are dominated by the voxels with the highest HI density. The reason for this is that 21cm IM is sensitive to the mass-weighted HI, rather than the volume-weighted HI of, e.g., the Ly$\alpha$ forest. Unfortunately, those dense voxels are rare in the simulation.
%A bulk of the cosmological information relevant for summary statistics comes from the high mass halos (or equivalently the densest voxels), which are encountered rarely in the simulation volume. 
For instance, there are $\sim 10^5$ halos with $M_\textup{halo}\geq10^{10} h^{-1} M_\odot$ in our dataset, which translates in a very small subset ($\sim$1 in 10$^3$) of voxels of our training sample having a non-negligible HI density (see the distribution of the voxel \HI masses in Figure~\ref{fig:HistVoid}).
%Most of the information relevant for our summary statistics comes from a very small subset ($\sim$1 in 10$^3$) of our training sample which have halos .
Because of such \emph{sparsity} in our data distribution,
our model could easily achieve a high accuracy by predicting the low mass \HI voxels, ignoring the high mass \HI voxels; this makes our model harder to train. 

%One has to therefore add more weight to the high mass voxels in the loss function which we will describe later.
Modeling the fields in Lagrangian space rather than Eulerian space can in principle reduce the sparsity problem.
This is because in lagrangian space we use the displacement field, which is distributed over a larger region of space, as compared to the density field, which is largely concentrated inside the halo boundaries \citep{HeLiFen1907}. However, modeling the \HI field is not possible in lagrangian space because, unlike DM, the number of gas particles are not fixed in the simulation\footnote{Gas particles can form stars, that later may explode as supernovae and may form black holes.}.

A similar sparsity challenge exists when predicting the galaxy positions from a 3D DM field.
Some of the previous neural network based studies have tried to overcome this challenge by using a combination of two neural networks (two-phase model) \citep{ ZhaWanZha1902, YipZha19, ModFenSel18}: the first phase predicts the halo/galaxy position and the second phase predicts the mass of the halo/ number of galaxies. However, \HI, unlike galaxies, is scattered over a wide volume of the Universe, not just at the centers of large DM halos. Ignoring the low mass voxels would remove the Ly$\alpha$ forest from our data which is not ideal as the Ly$\alpha$ forest is a powerful cosmological probe by itself, and its contribution to the 21cm signal at high-redshift becomes more important \citepalias{VilGenCas1810}. %predict that $\sim 10$\% of \HI lies outside halos at $z=$5). %(although we find that this does not affect the summary statistics).

We therefore implement a different kind of two-phase model:  
%Both the phases of our architecture predict the \HI mass but for voxels with different \HI mass; 
the first (second) network is geared towards predicting the low mass (high mass) \HI voxels.
We have used the same U-Net architecture shown in Figure~\ref{fig:Architecture} for each of the two phases of our model. We provide further details of the two-phase model in Appendix~\ref{apx:details}.

\subsection{Training the network}
\label{sec:training}
As the memory of GPUs is limited, we split the 75$\Mpc$ TNG100 volume into smaller sub-boxes. We train the U-Net to predict \HI boxes of side length $\sim$1.17 h$^{-1}$ Mpc and containing $8^3$ voxels using input DM boxes of side length $\sim$2.34 h$^{-1}$ Mpc and $64^3$ voxels. As illustrated in Figure~\ref{fig:Architecture}, we want to predict the \HI in the sub-box residing at the center of a larger DM input. The motivation of using the larger DM box is to account for environmental information for voxels near the boundaries of the HI volume. %\citetalias{VilGenCas1810} showed that a smaller halo residing in the environment of a larger halo is expected to contain less \HI on average than the smaller halo which is isolated in the field.

We obtain $\sim 2.5 \times 10^5$ non-overlapping \HI sub-boxes when splitting the TNG100-1 volume. We divide these sub-boxes in three chunks: $\sim$ 60\% of the sub-boxes are used for training the network, 12.5\% for validation and 27.5\% for testing the network. 
%We have reserved a large fraction of our total data for testing the U-Net.
We have constructed the test set such that it comprises of all the sub-boxes which correspond to a larger box of side-length $\sim$ 48 $\Mpc$. %the test set will be used for comparing the results of our network in Section~\ref{sec:results}.
The total number of trainable parameters in the U-Net shown in Figure~\ref{fig:Architecture} is $2.1\times10^7$, which, although seems gigantic, can be optimized using the technique of automatic differentiation (gradient descent). The gradients are calculated based on the 
%Table~\ref{tab:config} presents the details of the Hyper-parameters used in our network. 
following loss function
\beq
\mathbb{L}=\sum_i^{\textup{Voxels}} (\tilde{\delta}^\textup{pred}_i-\tilde{\delta}^\textup{Illustris}_i)^2 \times \exp[(\tilde{\delta}^\textup{Illustris}_i -\beta)^\alpha]
\label{eq:Loss}
\eeq
where $\tilde{\delta}$ is the scaled \HI density field from Eq.~(\ref{eq:DataTransform}). Notice that our loss function is different from the traditional mean square error (MSE) loss; we use the additional hyperparameters $\alpha$ and $\beta$ to add more weight to the high mass voxels in order to alleviate the aforementioned sparsity problem\footnote{It is worth noting that the modified loss function is well behaved and does not diverge for physically plausible values of the \HI field (in our sample, $\tilde{\delta}^\textup{Illustris}_\textup{i}\leq 2.93$).}. We find that $\alpha=2$ and $\beta=0.7$ give the best results. We provide further details on the training of the network in  Appendix~\ref{apx:details}. We have also made use of the techniques of data augmentation and imbalanced sampling to tackle the sparsity problem and we provide further details in Appendix~\ref{apx:augmentation}.

%---------------------------------------------------------------
\section{Benchmark model: Halo Occupation Distribution (HOD)}\label{sec:HOD}

We will compare the results of our neural network against a benchmark model, that we describe in detail in this section.

There have been multiple recent attempts at developing a halo model for the abundance and spatial distribution of \HI \citep{VilVieDat14,EmaPaco_17,VilGenCas1810,PadChoRef16,Spi20}.
The main idea behind those models is that most of the \HI mass in the post-reionization era is inside halos: more than 99\% at $z<0.2$ (the fraction decreases to 88\% at $z=0.5$) \citepalias{VilGenCas1810}. We will use the halo model of \citetalias{VilGenCas1810} as a benchmark to compare the performance of the neural network. We briefly describe their model here and refer the reader to \citetalias{VilGenCas1810} for further details.
The first step to produce 3D \HI density fields through the HOD method consists in running a DM-only simulation and identifying halos: saving their positions, masses and radii.
A DM halo of FOF (friends-of-friends) mass $M$ is then assigned an \HI mass:
\beq
M_\textup{HI} (M,z) = M_0 \left(\frac{M}{M_\textup{min}}\right)^\alpha \exp [-(M_\textup{min}/M)^{0.35}]
\label{eq:HOD}\eeq
where $M_0$ is a normalization factor, $\alpha$ is the power-law slope, $M_\textup{min}$ is the characteristic minimum mass of halos that host \HII. These three parameters were fitted to reproduce the results of the TNG100-1 simulation by \citetalias{VilGenCas1810}, and their best-fit values at $z=1$ are: $M_0 = 1.5 \times 10^{10} h^{-1} M_\odot$, $M_\textup{min}=  6 \times 10^{11} h^{-1} M_\odot$ and $\alpha=0.53$.
Given the total \HI mass inside a halo, the HOD will provide its spatial distribution within the halo, i.e. its \HI density profile. In small halos, \HI is typically localized in their inner regions. For groups and galaxy clusters, the central region of the halo is typically \HI poor, due to the action of processes such as AGN feedback, ram-pressure and tidal stripping. Therefore \citetalias{VilGenCas1810} fitted a simple power law with an exponential cutoff on small scales given by 
\beq
\rho_\textup{HI}(r) = \frac{\rho_0}{r^{\alpha_*}} \exp (-r_0/r)
\label{eq:1haloterm}\eeq
for the \HI density profile. We implement this density profile by assigning 200 particles to each halo following the density profile in Eq.~(\ref{eq:1haloterm}). For simplicity, we adopt $\alpha_*=3$ and $r_0=1$ $h^{-1}$ kpc for all halo masses \citepalias{VilGenCas1810}. Note that if we do not include the one halo term, the \HI power spectrum becomes dominated by shot noise at $k\sim1$ h/Mpc \citepalias{VilGenCas1810}. It is important to note that the HOD parameters have not been tuned to match the summary statistics like the power spectrum (as is typically done in the case of galaxy survey data analysis), rather the HOD is fitted to the $M_\textup{HI}$ vs $M_\textup{HOD}$ plot from the TNG data (see Fig.~3 of \citetalias{VilGenCas1810}). Our goal here is to test the HOD model and compare its performance with our neural network approach.

While the HOD performs well on high density \HI regions in the Universe like the Damped Lyman Absorbers, it is expected to perform poorly for systems with low \HI density like the Ly$\alpha$-forest. Other drawbacks of HOD is that it relies on simplistic parameterizations like in Eq.~(\ref{eq:HOD}), and assumes a spherical distribution of \HI within halos. More importantly,
the only information used for predicting $M_\textup{HI}$ is the halo mass and all other properties like the environment of the halo and its concentration are ignored; we will return to this point in section \ref{sec:discussion}.

%-------------------------------------------------------------------------------
\section{Results}\label{sec:results}
\begin{figure*}
    \centering
        \includegraphics[scale=0.55,keepaspectratio=true]{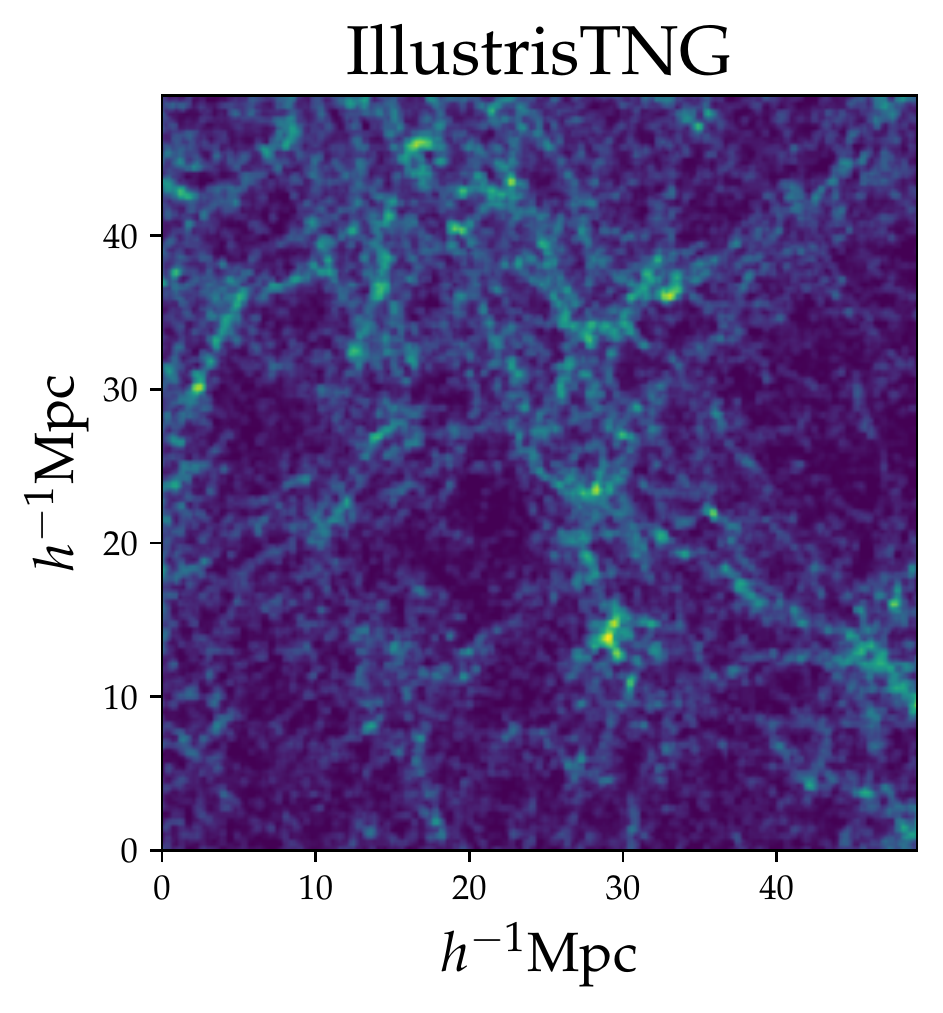}
        \includegraphics[scale=0.55,keepaspectratio=true]{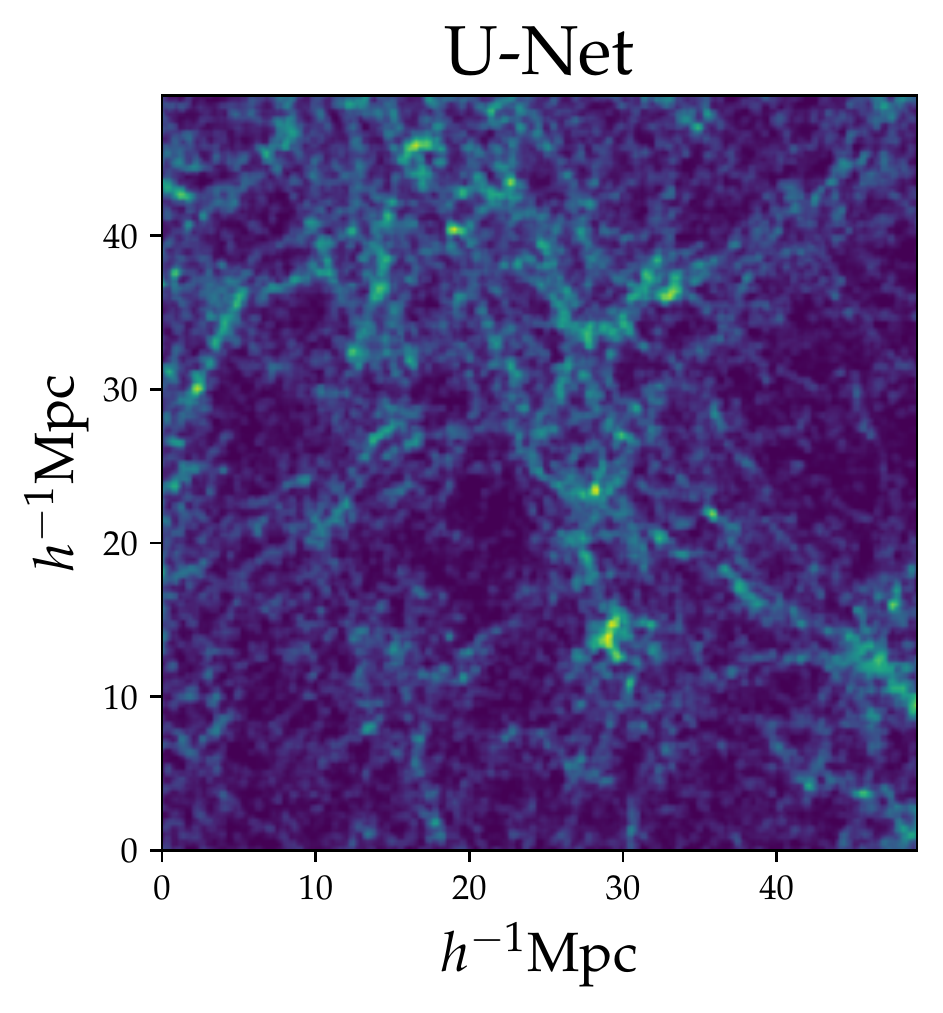}
        \includegraphics[scale=0.55,keepaspectratio=true]{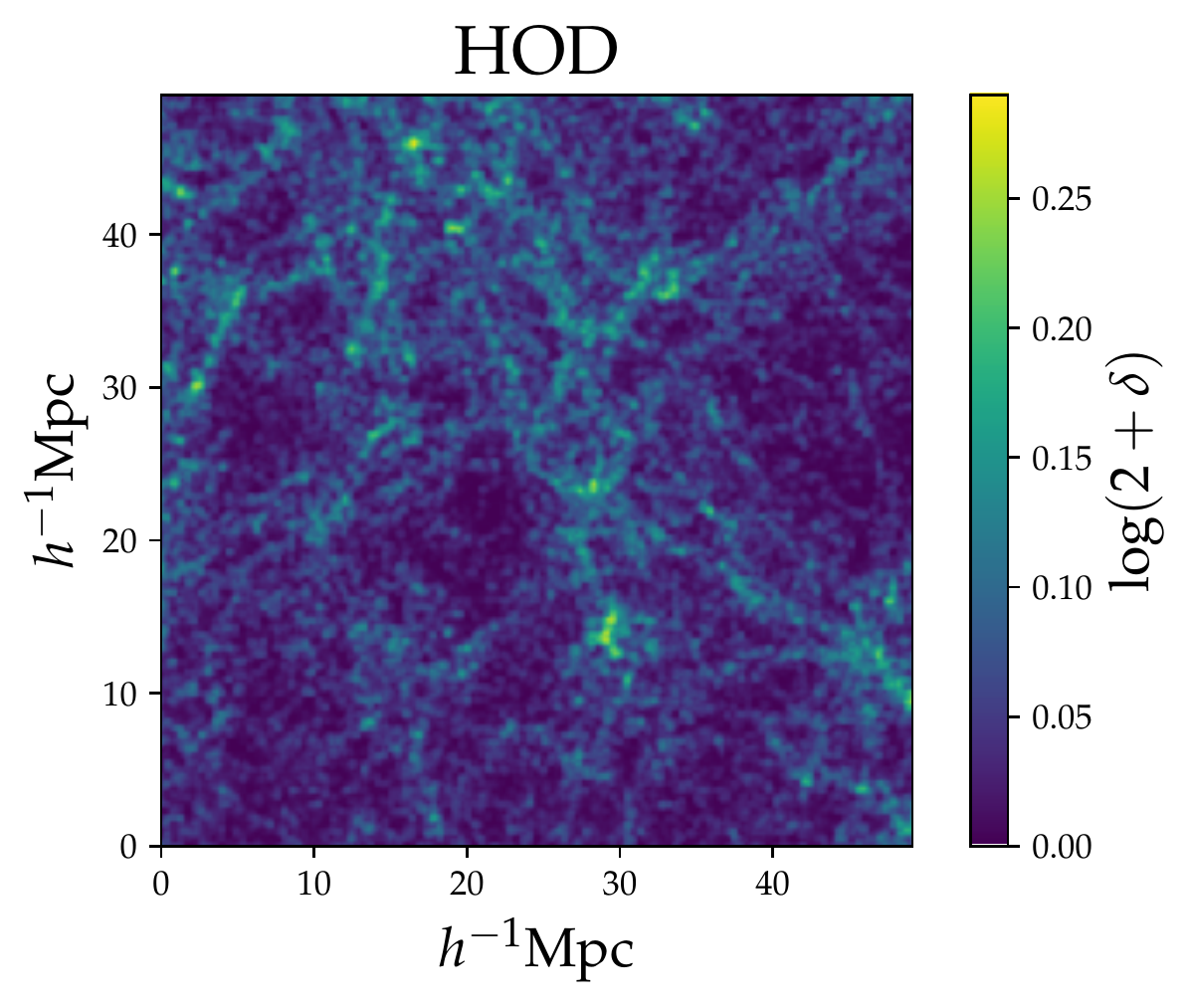}
        \includegraphics[scale=0.55,keepaspectratio=true]{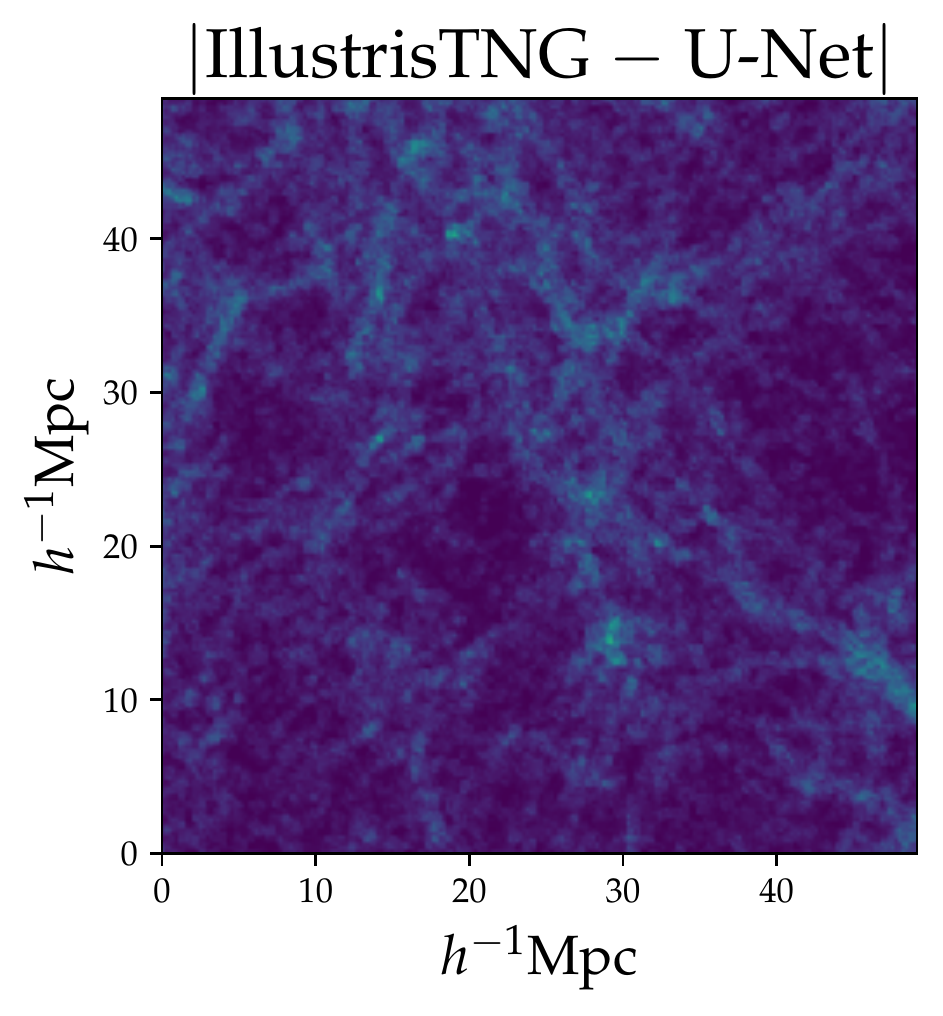}
        \includegraphics[scale=0.55,keepaspectratio=true]{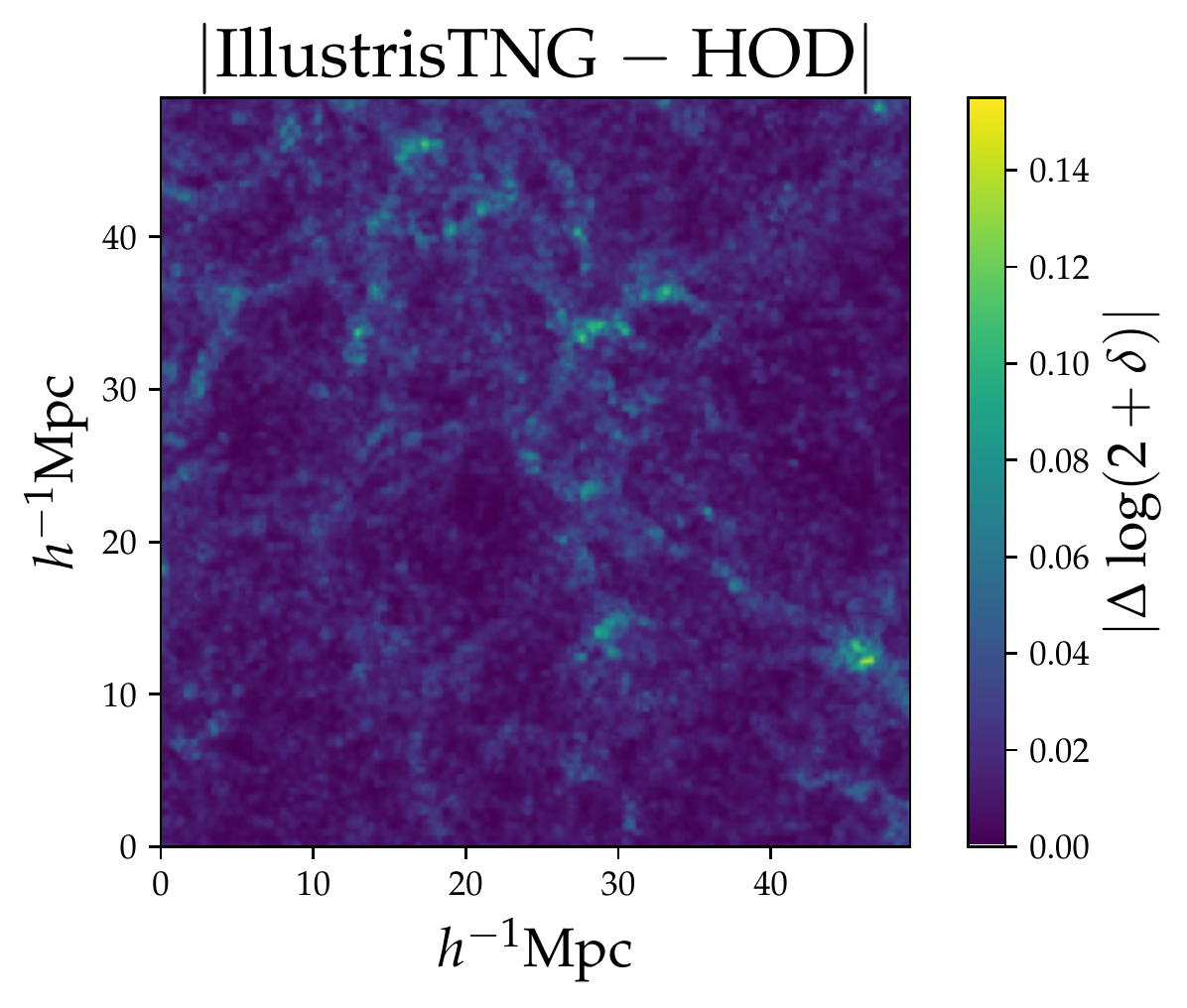}
\caption{The top row shows the projected \HI density field at $z=1$ from a region of (48 Mpc/h)$^3$ for the labelled cases. The bottom left (right) panel shows the residuals between the IllustrisTNG simulation and the \HI fields obtained from the U-Net and HOD method. The color scale is anchored for each row. The residuals for U-Net are smaller than those for HOD in areas of high \HI density.}
    \label{fig:2D}
\end{figure*}

In this section we present the results our the neural network and its comparison with the HOD model.

We have reserved the sub-cubes corresponding to a larger cube of side 48 $\Mpc$ in the IllustrisTNG simulation volume for testing our network. Once our network is trained, we concatenate the generated \HI field corresponding to all the sub-cubes and use the larger cube to compare the summary statistics.
A possible concern could be that the stacking of the individual \HI boxes to make a larger box can lead to spurious edge effects in the summary statistics, but we have checked that such effects are negligible, see Appendix~\ref{apx:edge_effects} for further details.
 
We first show a visual comparison of the network output in Figure~\ref{fig:2D}. In the bottom panels we have averaged over the absolute values of the differences in the fields along the projected axis. We now discuss multiple summary statistics and find that our network outperforms HOD up to the non-linear scales $k \lesssim 1$ h/Mpc in all the statistics.

\subsection{HI Power Spectrum}
\begin{figure*}
\centering
\includegraphics[scale=0.58,keepaspectratio=true]{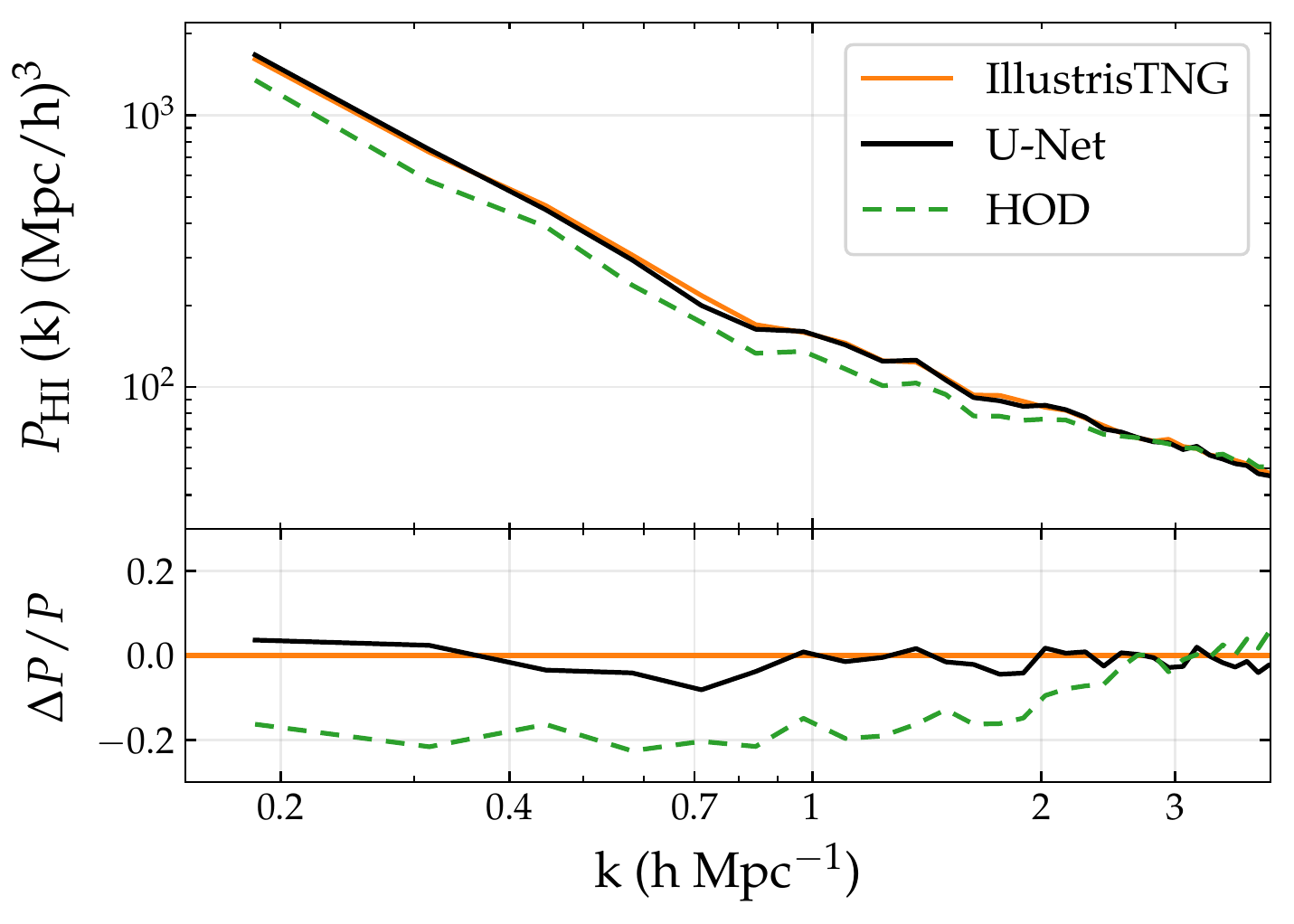}
\includegraphics[scale=0.58,keepaspectratio=true]{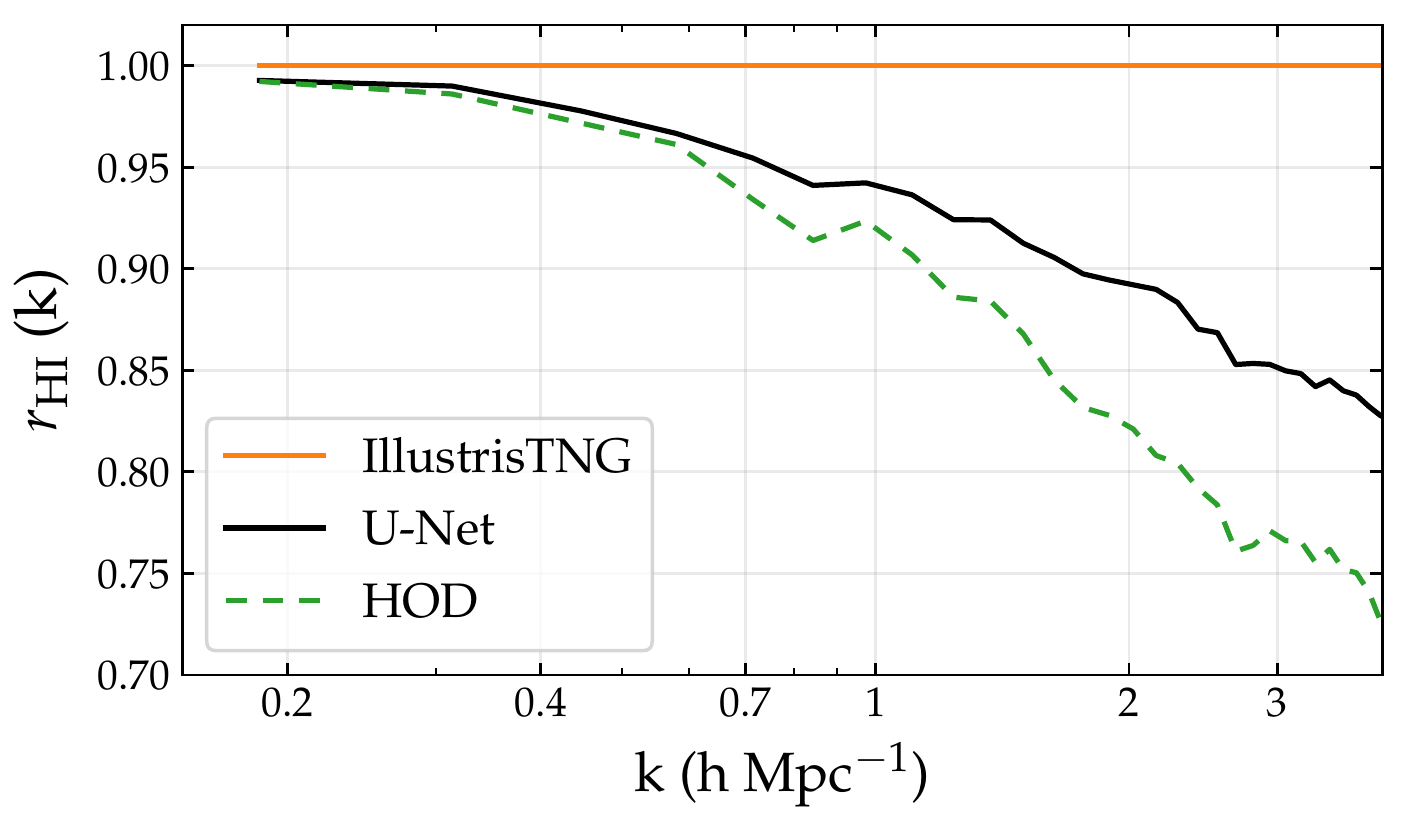}
\caption{Comparison between the \HI power spectrum [left] and the cross-correlation coefficient, defined as $r_\mathrm{(A)} = P_\mathrm{A-Illustris}/ \sqrt{P_\mathrm{A} P_\textup{Illustris}}$, [right] for the IllustrisTNG simulation (orange), the HOD (green) and the neural network (black). We can see that the network outperforms the HOD for both statistics on all scales.}
    \label{fig:Pow}
\end{figure*}

The most widely used summary statistic in cosmology is the power spectrum, which is the Fourier transform of the two-point correlation function. In 21cm IM, the quantity that is directly observed is the 21cm power spectrum, which is related to the HI power spectrum via
%One cannot directly measure the fluctuations in the \HI density in the Universe but instead one measures the fluctuations in the brightness temperature of the 21cm line.
%The power spectrum of \HI density field $P_\textup{HI}$ is related to the observable 21cm power spectrum as
\beq
P_\textup{21cm}(k) = \bar{T}^2_b\, P_\textup{HI}(k)\, ,
\label{eq:P21cm}\eeq
where $\bar{T}_b$ is the mean brightness temperature of the 21cm line at redshift $z$. Let us now compare the two terms in the RHS separately. The mean brightness temperature scales as
$T_b \propto \Omega_\textup{HI}$, where $\Omega_\textup{HI}$ is the ratio between the density of \HI at redshift $z$ and the Universe's critical density at $z=0$.
% \beq
% \bar{T}_b(z)=189 h \bigg(\frac{H_0(1+z)^2}{H(z)}\bigg) \Omega_\textup{HI}\ \textup{mK}.
% \eeq
The network and the HOD predicts values of $10^{4} \times \Omega_\textup{HI}(z=1)$ in our test set to be  5.77 and 6.43, respectively, while that value is 5.82 for the IllustrisTNG simulation.
%We find the values of $10^{4} \times \Omega_\textup{HI}(z=1)$ to be 5.82, 5.77, 6.43 for the case of IllustrisTNG, our network and HOD respectively.

The second term to compare in Eq. (\ref{eq:P21cm}) is $P_\textup{HI}(k)$, and we show the results from different methods in Figure~\ref{fig:Pow}. We find that the network is able to reproduce the HI power spectrum from the simulations up to a deviation of $\lesssim 5$\% all the way to non-linear scales $k \lesssim 1$ $\kMpc$, whereas the HOD deviates by $\lesssim 20$\% at low-$k$ although it becomes slightly more accurate at high-$k$.
The increase in the HOD accuracy at high-$k$ arises mainly due to the 1-halo term given in Eq.~(\ref{eq:1haloterm}). It is important however to note that the one-halo term in Eq.~(\ref{eq:1haloterm}) only aids in the accuracy of the power spectrum (amplitude of fluctuations) at high-$k$ but not does not accurately model the phases of the \HI fluctuations, which are relevant for the cross-correlation coefficient $r_\textup{HI}$ (compared in the right panel of Figure~\ref{fig:Pow}). \citetalias{VilGenCas1810} argued that one possible cause of the discrepancy in the HOD power spectrum at low-$k$ may be be due to the fact that they do not explicitly fit the total $\Omega_\textup{HI}$ in the simulation volume, which would change the \HI bias. However, even if the bias is changed, their $r_\textup{HI}$ should remain unchanged.
 
We do not show error bars arising from sample variance in the testing volume because the both the DM and the \HI fields from IllustrisTNG are evolved from the same initial conditions. We chose the particular range of $k$ in Figure~\ref{fig:Pow} because of the two following constraints: the low-$k$ limit is set by the largest mode in the test set (making an even larger test set is possible but at the expense of reduction in the training data), and the high-$k$ limit is conservatively set such that the scales affected by the smoothing of the \HI maps ($\sim$ 5 $\times$ 300 $\Kpc$) are removed (see figure~\ref{fig:Pow_k=10} for $P_\mathrm{HI}$ plotted till $k=10 \kMpc$).

It is worth noting that the accuracy of $P_{\rm HI}(k)$ is dependent on modeling of \HI in high-mass halos (see for e.g. Figure~\ref{fig:2-phase}).
As mentioned in Section~\ref{sec:DataSparsity}, the training of our network is challenging in the high-mass end, because of the sparsity associated with the the abundance of voxels in that regime.
Our results can be further improved if we train the U-Net on a simulation with a larger volume\footnote{We plan to train the U-Net on the Illustris TNG-300 sample (which has three times the volume of the TNG-100 sample) and expect to find even better agreement at the high-mass end of the PDF in Figure~\ref{fig:HistVoid}.}, which would have a larger number of high mass halos, or using a set of zoom-in simulations focused on galaxy clusters \citep{Thiele_2020}.

Let us now estimate, very crudely, the accuracy of the power spectrum model required for future 21cm surveys using some fisher forecasts in the literature. \cite{ObuCas18} showed the peak signal-to-noise ratio (S/N) could be $\sim 10$ for upcoming surveys like HIRAX and CHIME (note that S/N will be much larger for future surveys like SKA), and therefore the accuracy of the method to produce 21cm mocks should roughly be $\lesssim10\%$ for the power spectrum.

\subsection{Cross-correlation with galaxies and halos}
Large regions of future \HI surveys will overlap with regions sampled by galaxy spectroscopic surveys like DESI or Euclid.
One of the most important summary statistic in such overlapping regions is the \HI-galaxy cross-power $P_{\textup{HI} - \textup{Galaxy}}$. Using the cross-correlations will boost the S/N ratio and mitigate the effects of foregrounds (e.g. \cite{VilVieAlo15} predict the S/N for the cross-correlation between future 21cm surveys and Lyman-Break galaxies to be larger than 21cm auto-correlation by a factor of $\sim 10$ ($\sim$ 4) at large (small) scales; see also \cite{PadRef20,ModWhi21}).
Furthermore, unlike the auto-power spectrum of 21cm which is yet to be detected, there have already been multiple detections of the $P_{\textup{HI} - \textup{Galaxy}}$ signal \citep{ChaPen10,MasSwi13,AndLuc18}.

\begin{figure}
\centering
\includegraphics[scale=0.55,keepaspectratio=true]{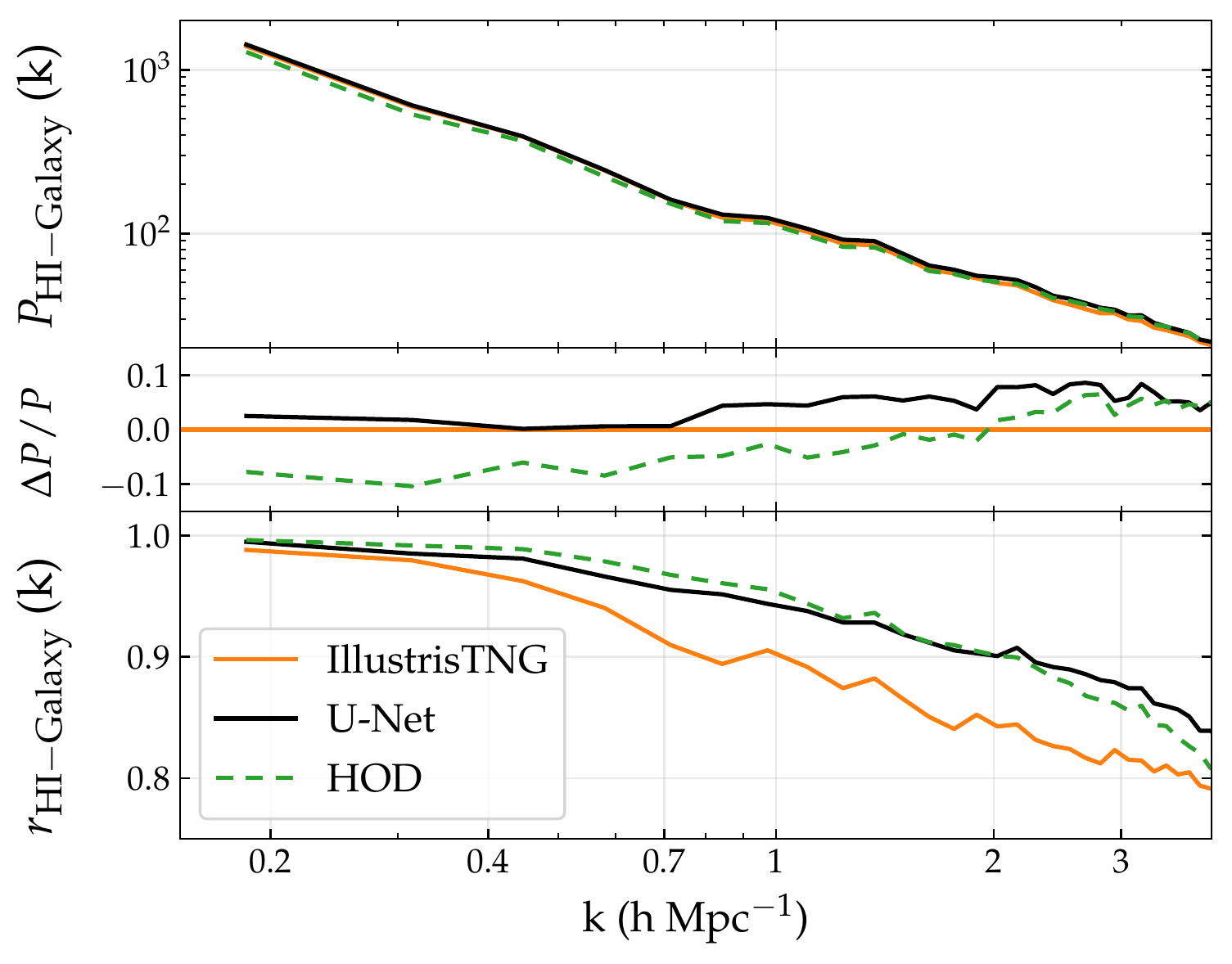}
\caption{Comparison of the \HI - galaxy cross power (upper panel), its relative error (middle), and the \HI - galaxy cross correlation coefficient (lower), which is defined as $r_\textup{HI - Galaxy} = P_{\textup{HI} - \textup{Galaxy}}/ \sqrt{P_{\textup{HI}} P_\textup{Galaxy}}$). The line labels are same as in Figure~\ref{fig:Pow}. The network outperforms HOD up to the non-linear scales  ($k \lesssim$ 1 $\kMpc$).}
    \label{fig:GalaxyCrossPow}
\end{figure}

In Figure~\ref{fig:GalaxyCrossPow}, we compare $P_{\textup{HI} - \textup{Galaxy}}$ from our approach.
%We therefore use the IllustrisTNG galaxy field to calculate the \HI-galaxy cross-power $P_{\textup{HI} - \textup{Galaxy}}$ and its cross-correlation coefficient in
%Figure~\ref{fig:GalaxyCrossPow}. 
We have included all the galaxies in the TNG100-1 sample with the stellar mass M$_*> 10^{10}$ M$_\odot$
(which roughly corresponds to a number density of $n=10^{-3}$ h$^3$/Mpc$^3$)
for this measurement. Similar to the $P_\textup{HI}(k)$ in Figure~\ref{fig:Pow}, the $P_{\textup{HI} - \textup{Galaxy}}$ exhibits a bias for the HOD at low-$k$.

\begin{figure*}
\centering
\includegraphics[scale=0.58,keepaspectratio=true]{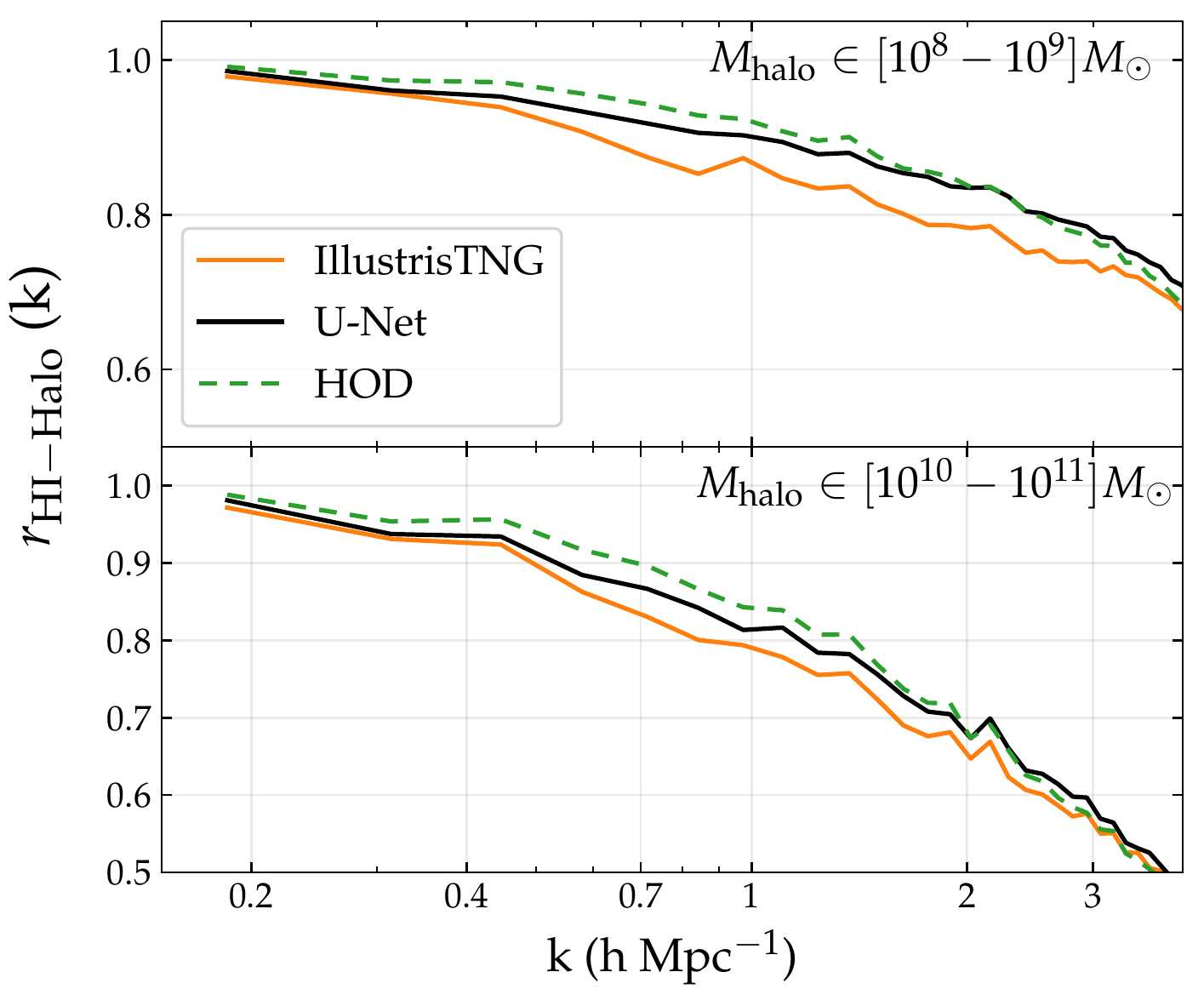}
\includegraphics[scale=0.58,keepaspectratio=true]{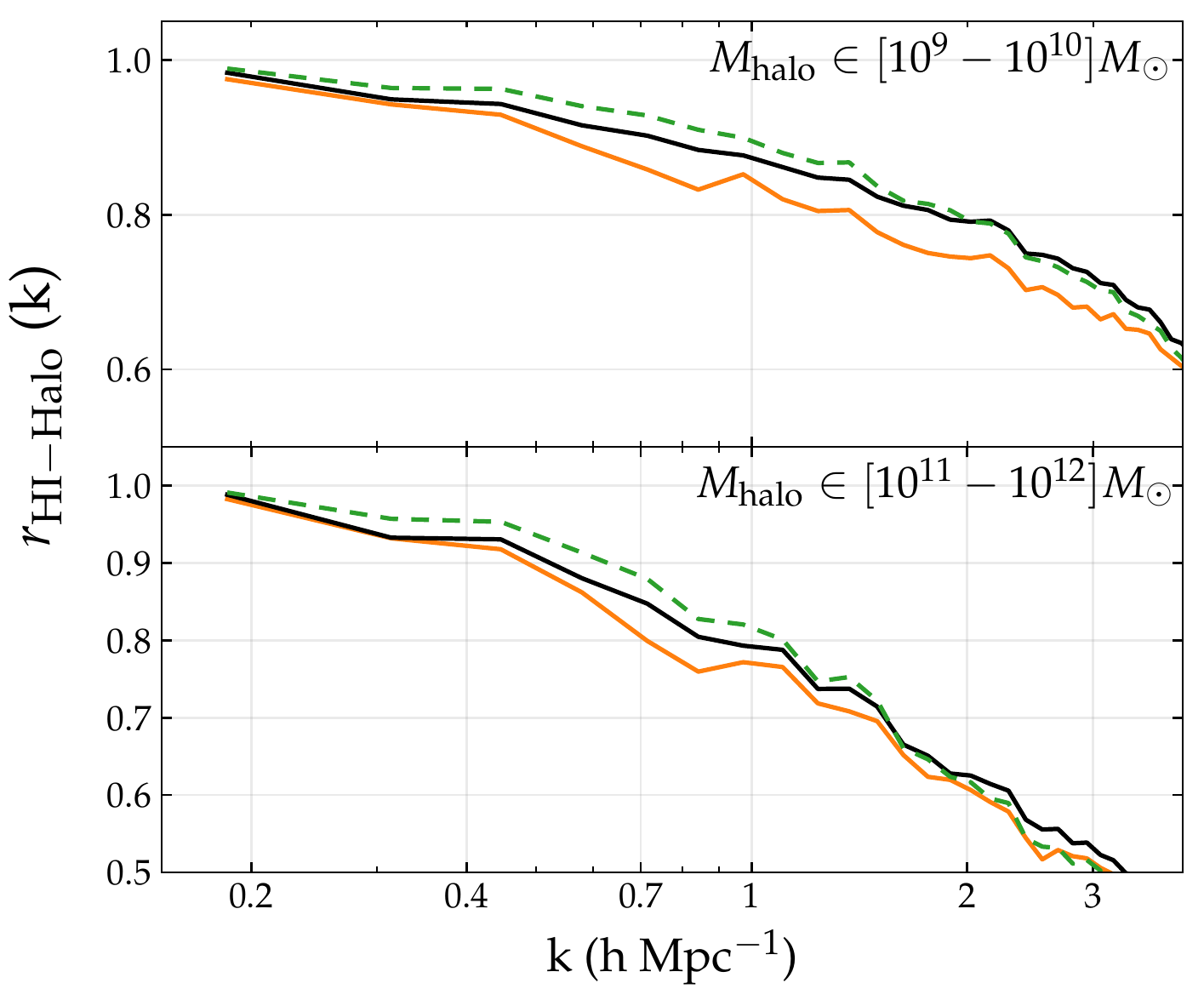}
\caption{\HI - halo cross correlation coefficient ($r_\textup{HI - Halo} = P_{\textup{HI} - \textup{Halo}}/ \sqrt{P_{\textup{HI}} P_\textup{Halo}}$) for multiple halo mass bins. Line labels are same as in Fig. \ref{fig:Pow}. The network consistently outperforms the HOD up to non-linear scales ($k\lesssim  1 \kMpc$).}
\label{fig:HaloCross}
\end{figure*}

Another important statistic along the same lines is cross-correlation coefficient of \HI with halos, which is more sensitive to the way \HI mass is distributed across halos and to the one-halo term. We see in Figure~\ref{fig:HaloCross} that the U-Net outperforms the HOD up to $k\lesssim$ 1 h/Mpc for all bins of halo masses that we have considered.
It is worthwhile to note that if we extend Figures~\ref{fig:GalaxyCrossPow} and \ref{fig:HaloCross} for $k>4 \kMpc$, 
the green HOD curve diverges further away from the orange IllustrisTNG curve. It is therefore likely a coincidence that the HOD outperforms the U-Net for $1\lesssim k\lesssim 4 \kMpc$, as the green curve is merely crossing the orange curve to the other side.
However, we do not show the scales $k>4 \kMpc$ as they are affected by the smoothing of \HI maps and a high-resolution test has to be performed to be definitive.

\subsection{HI Bispectrum}

If all the fluctuations in the Universe were perfectly Gaussian, the field could be perfectly characterized by its two-point correlation function or its power spectrum. However, even if the primordial fluctuations were Gaussian, late-time gravitational clustering causes significant leakage of Gaussian information in the non-linear regime \citep{ScoZalHui9912,TakJai04,WadSco19,Quijote}. To recover this information, the lowest order statistic that one needs to compute in Fourier space is the bispectrum. Unlike the power spectrum, the bispectrum is sensitive to the shape of the structures generated by gravitational instability and has the promise to break important degeneracies in the bias and cosmological parameters \citep{Sco00,SefCro06,HahVilCas20,HahVil20,YanPor19,ChuIva19,KamSle20}. The post-reionization 21cm signal is expected to have significant information in the non-linear regime and there have been recent attempts at theoretical modeling of the \HI bispectrum \citep{SarMajBha19}. One of the toughest parts in a bispectrum analysis is calculation of the error due to cosmic variance and a fast technique to generate mock \HI fields is therefore essential.  

\begin{figure*}
\centering
\includegraphics[scale=0.58,keepaspectratio=true]{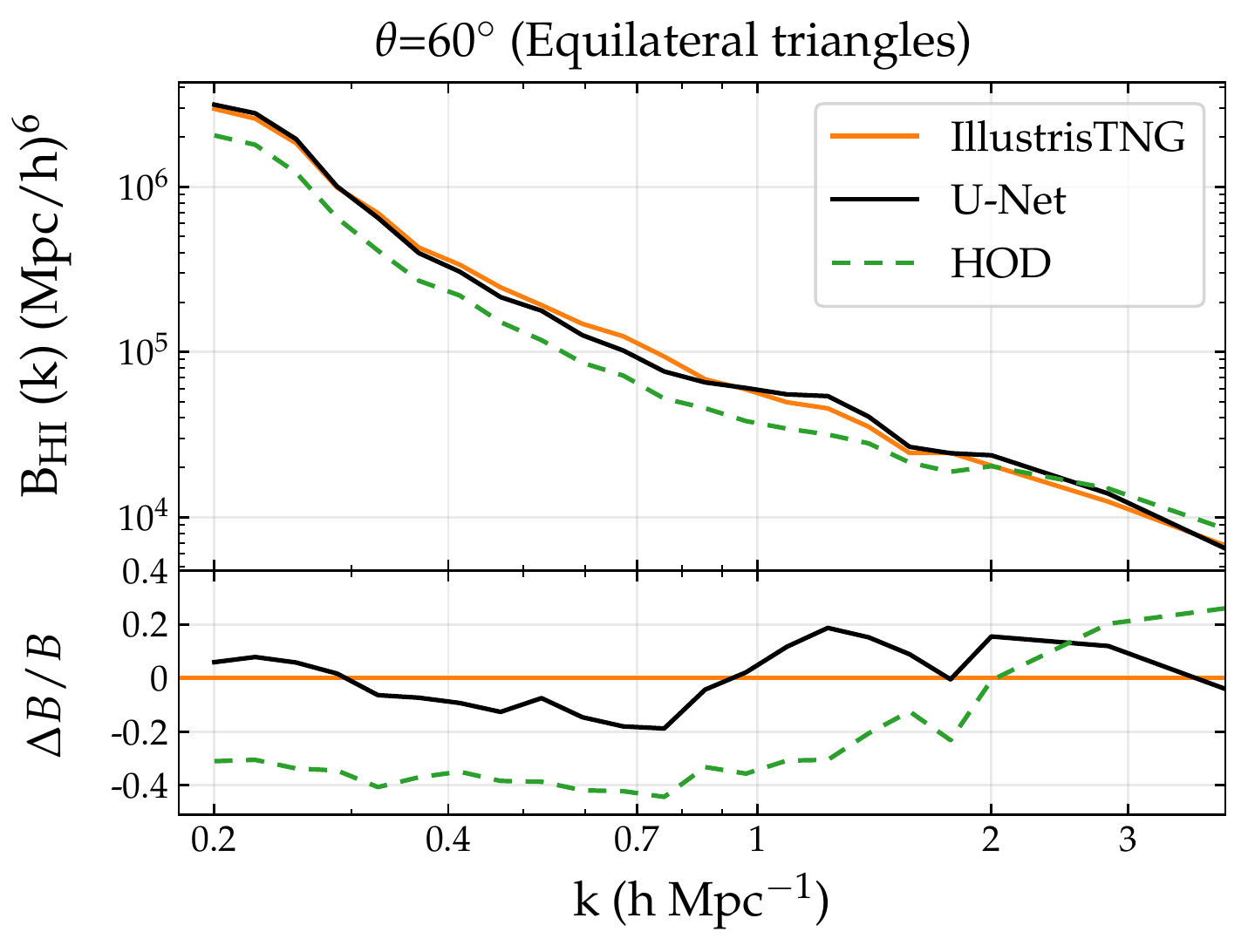}
\includegraphics[scale=0.58,keepaspectratio=true]{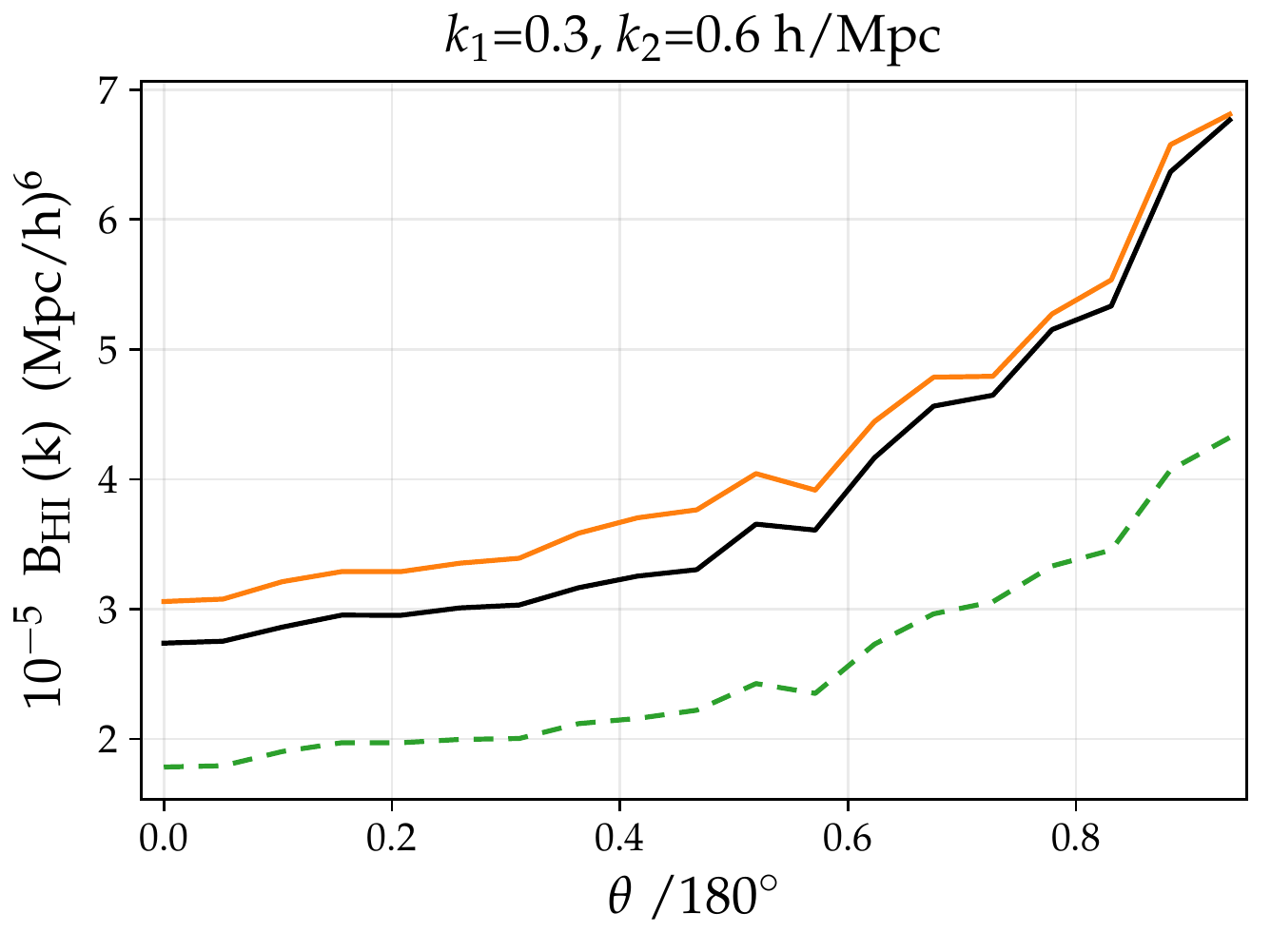}
\caption{The left panel shows the comparison of the bispectrum as a function of side length for the equilateral triangle configuration. The right panel shows the bispectrum when the angle between two particular sides of the triangle is varied keeping the length of those sides $k_1, k_2$ fixed. Line labels are same as Figure~\ref{fig:Pow}. Note that the network is able to reproduce the bispectrum better than HOD for nearly all triangle shapes and sizes.}
\label{fig:Bisp}
\end{figure*}

In Figure~\ref{fig:Bisp} we show the HI bispectrum of the IllustrisTNG simulation, together with the predictions of the HOD and neural network. We show two particular cases (which are representative of all possible triangle configurations): the first case is for equilateral triangles with different side-lengths and the second case is for triangles with a varying angle between two of its sides whose lengths are kept fixed. %\Paco{You need to expand this a bit more. Say that the first case if for equilateral triangles....etc}.
We checked that the results are similar for other triangle configurations.
The residuals of the bispectrum of equilateral triangles from the network compared to the target is $\lesssim 20$ \% for all scales we compared ($0.2 <k<4 \kMpc$), and the residuals for HOD on other hand are $\lesssim 45$ \%.

\subsection{1-D Probability distribution function}
\label{sec:1dPDF}
The upcoming \HI surveys will observe systems ranging from low column densities (Ly$\alpha$ forest) to very high column densities (Damped Lyman Absorbers); the PDF of \HI is a statistic which is sensitive to its distribution over the wide density range. The \HI PDF can also be used for probing non-Gaussian information and for constraining luminosity functions \citep{Bre17,LeiUhl19}. We show in Figure~\ref{fig:HistVoid} the comparison of the U-Net prediction over four orders of magnitude of \HI voxel masses. The comparison is difficult in the high-mass end due to sample variance: that regime is dominated by very massive halos, which are rare. Note that there is a slight discrepancy in the HOD approach for low \HI mass voxels as it does not take into account filaments and also misses very low-mass halos which are below the simulation resolution threshold.
%We have also checked that smoothing the density field over a particular radius and comparing the PDF.

\subsection{Abundance of HI voids}
\begin{figure*}
\centering
\includegraphics[scale=0.58,keepaspectratio=true]{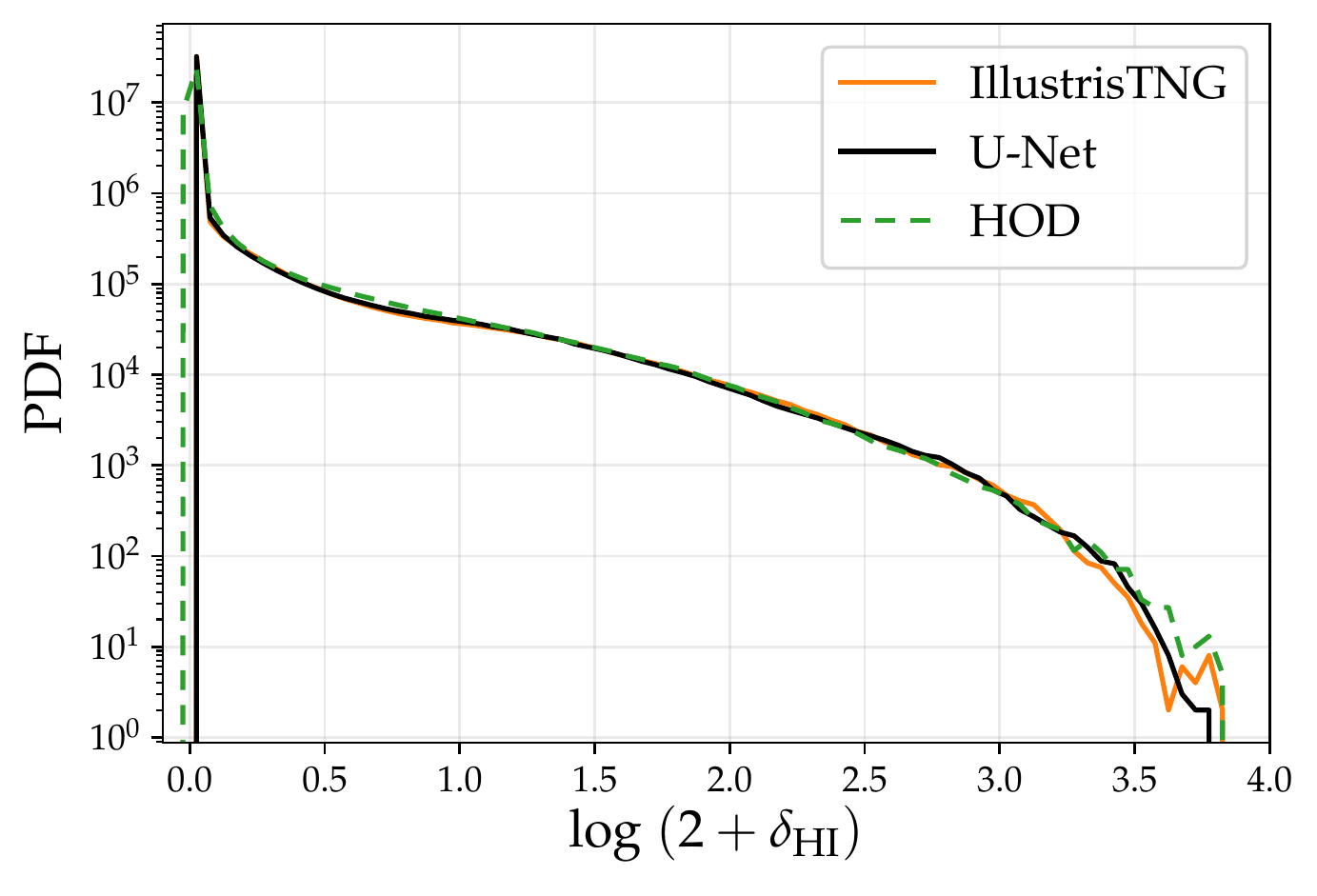}
\includegraphics[scale=0.58,keepaspectratio=true]{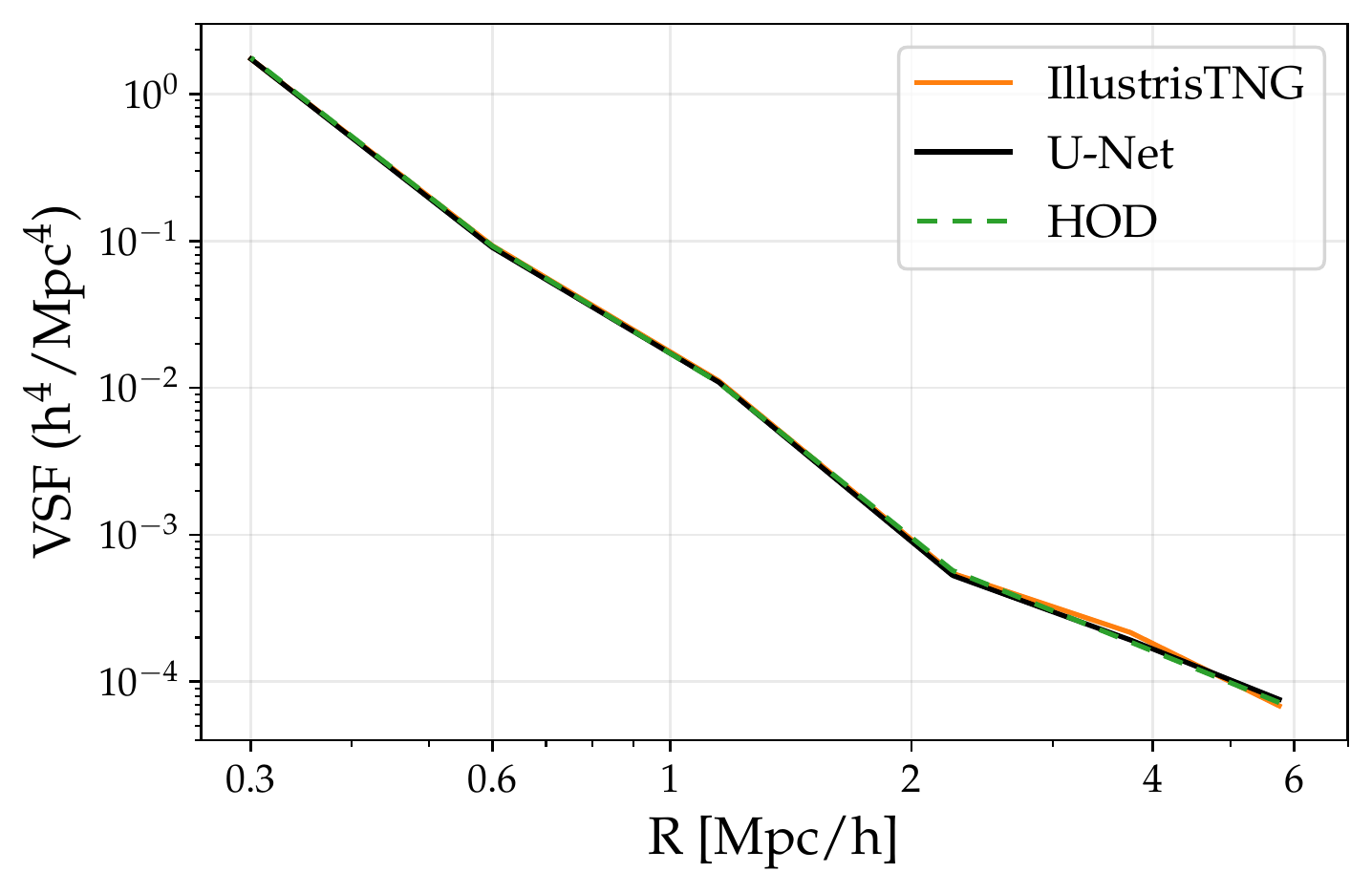}
\caption{Comparison of the 1D \HI PDF (left) and the void size function (right). Line labels are same as Figure~\ref{fig:Pow}. As there are very few high mass halos in our test set volume, the high-mass tail of the PDF is dominated by sampling noise. Both our network and HOD reproduce the above statistics to a good accuracy.}
    \label{fig:HistVoid}
\end{figure*}

Voids are the most underdense regions of the Universe. 
In Figure~\ref{fig:HistVoid} we show the void size function (VSF) of the HI field, which is defined as the number density of HI voids as a function of radius. We have used the algorithm described in \cite{BanDal16} to identify voids. The VSF is an important statistic as it contains complementary information to the one from traditional clustering observables.

%rest of the statistical properties that we have discussed in this section.
%\subsection{Additional tests}
%Make a deeper net

\section{Discussion}
\label{sec:discussion}

Let us now briefly compare our method to some other neural network approaches for emulating hydrodynamic simulations.
One difference from \cite{ZamOkaVil1904,ZhaWanZha1902,YipZha19,TroFer19} is that our method can produce a hydrodynamic field of any large size (and is not restricted to the size of the input simulation box used for training the network).
\cite{ZamOkaVil1904} also focuses on modeling \HI and used a generative adversarial network (GAN) to generate 3D samples of the \HI field at redshift $z=5$ on very small scales: between 35 $\Kpc$ and 2.34 $\Mpc$. However, their method cannot model the \HI fluctuations on large scales, which are relevant for 21cm experiments. Our method, on the other hand, can model \HI on all scales larger than $0.3$ $\Mpc$.

Our network takes $\sim 1.8$ hours to generate a \HI box of side 100 $\Mpc$ from a given DM box on a single GPU (for comparison, the IllustrisTNG simulation takes tens of millions of CPU hours for an equivalent volume \citep{NelPilAnn19}). Our method is therefore capable of making Gigaparsec volume mock \HI fields.

It is worth mentioning some of the caveats of using deep neural networks to make mock cosmological simulations. We have trained our model to emulate a particular IllustrisTNG simulation which has fixed values of parameters for cosmology and for various baryonic feedback prescriptions. The current model also uses a high-resolution input DM field. One should be very careful with extrapolating any machine learning model beyond the range of data that it has been trained on \citep{PfeBre19}. It is not obvious if, without recalibration, our model can emulate a simulation with a different cosmology or baryonic feedback prescription or can work with a lower resolution input DM field. However,
on the upside, our model only takes a couple of days to training to emulate a given simulation. Our technique is flexible and can quickly learn to emulate future hydrodynamic simulations which will be better than the current ones because of better technology and more observational data.

Let us discuss one interesting direction to be explored in future work.
Studies have shown that the distribution of baryons inside the halos are affected by its history (for e.g. the halo formation time \citep{JiaVan17}). Because the U-Net is very flexible on the dimensionality of the input field, we could take into account the halo history information by including multiple DM snapshots at different redshifts as input to our U-Net. This way the U-Net would be able to approximate a function $f$ of an even more general form that the one presented in Equation~(\ref{eq:f2}):
\begin{equation}
    \rho_{\rm HI}(\x,t)=f(\rho_{\rm m}(\x,t), \rho_{\rm m}(\x',t'))
\label{eq:f3}\end{equation}
which implies that to predict the \HI field at a particular point in space and time $(\x,t)$, the information should arise not only from the spatial vicinity $\x'$, but also from its time evolution $t'\leq t$.

%------------------------------------------------------------
\section{Conclusions}
\label{sec:conclusions}
Multiple upcoming radio telescopes such as CHIME, HIRAX, and SKA will be able to map the 21 emission from cosmic HI in the post-reionization Universe.
Mock \HI fields spanning Gigaparsec volumes are needed to provide theory predictions in the non-linear regime, to compute covariance matrices and to evaluate the effect of observational systematics like foregrounds, among many other things.

We use a deep convolutional neural network to find the mapping between the 3D fields of DM (from an N-body simulation) to \HI (from the IllustrisTNG hydrodynamic simulation). We compared the results of our network against a state-of-the-art HOD benchmark. We show that the neural network outperforms the results of the HOD in all the summary statistics considered: power spectrum, cross-correlation coefficient, bispectrum, PDF, and void size function. While the HOD method neglects any environmental dependence on the abundance of HI inside halos, our neural network can capture any underlying pattern present. We show in our more recent paper that there is indeed a significant effect of the environment of the halo on its \HI content, and we model it using more interpretable machine learning techniques like symbolic regression \citep{WadVilHo20inprep}. 

This study focuses on modeling \HI in real space and we will address the modeling in redshift space in a future work. We have used a DM field from a high resolution $N$-body simulation as an input in this analysis and we plan to explore whether the U-Net technique can work on lower resolution $N$-body simulations or other approximate gravity-only simulations.
Although we have focused on \HI in this paper, we anticipate neural networks to be able to produce mocks for other line intensity mapping surveys by emulating expensive hydrodynamic simulations.

\vspace{-1cm}
\acknowledgements{We thank Gabriella Contardo, David Spergel, Roman Scoccimarro and Leander Thiele for fruitful discussions. We are especially grateful to Yin Li for many enlightening discussions. We also thank Chirag Modi, Yin Li and especially an anonymous referee for useful comments on the manuscript. FVN acknowledges funding from the WFIRST program through NNG26PJ30C and NNN12AA01C. The work of SH is supported by Center for Computational Astrophysics of the Flatiron Institute in New York City. The Flatiron Institute is supported by the Simons Foundation. This work was also supported in part through the NYU IT High Performance Computing resources.
%The source code of our Pytorch implementation will be made available on Github \footnote{\url{https://github.com/JayWadekar/DMtoHI}}.
The IllustrisTNG data is publicly available at \url{https://www.tng-project.org/data/}, while the scripts used in this project are readily available upon request.
We have made use of the Pylians3 libraries (\url{https://github.com/franciscovillaescusa/Pylians3}) to carry out the analysis of the simulations. }

\begin{appendix}

\section{Details and Methods}
\label{apx:details}

\subsection{Details of network architecture}
We had presented our network architecture in Figure~\ref{fig:Architecture} and we discuss its details in this section. Let us first discuss details of the input and the output of our network. The input DM box has a physical side length $2.34 \Mpc$ with a grid
resolution of $\sim 35 \Kpc$ and therefore has $64^3$ grid cells. As discussed in section~\ref{sec:DataProcessing}, the output \HI box has a lower resolution (physical side length of $1.17 \Mpc$ with $8^3$ grid cells).  A U-Net typically consists of a contracting path and an expansive path of nearly equal lengths.
In our case, due to lower dimensionality of the output as compared to the input, the expansive path is relatively much shorter.

\begin{figure*}
\centering
\includegraphics[scale=0.58,keepaspectratio=true]{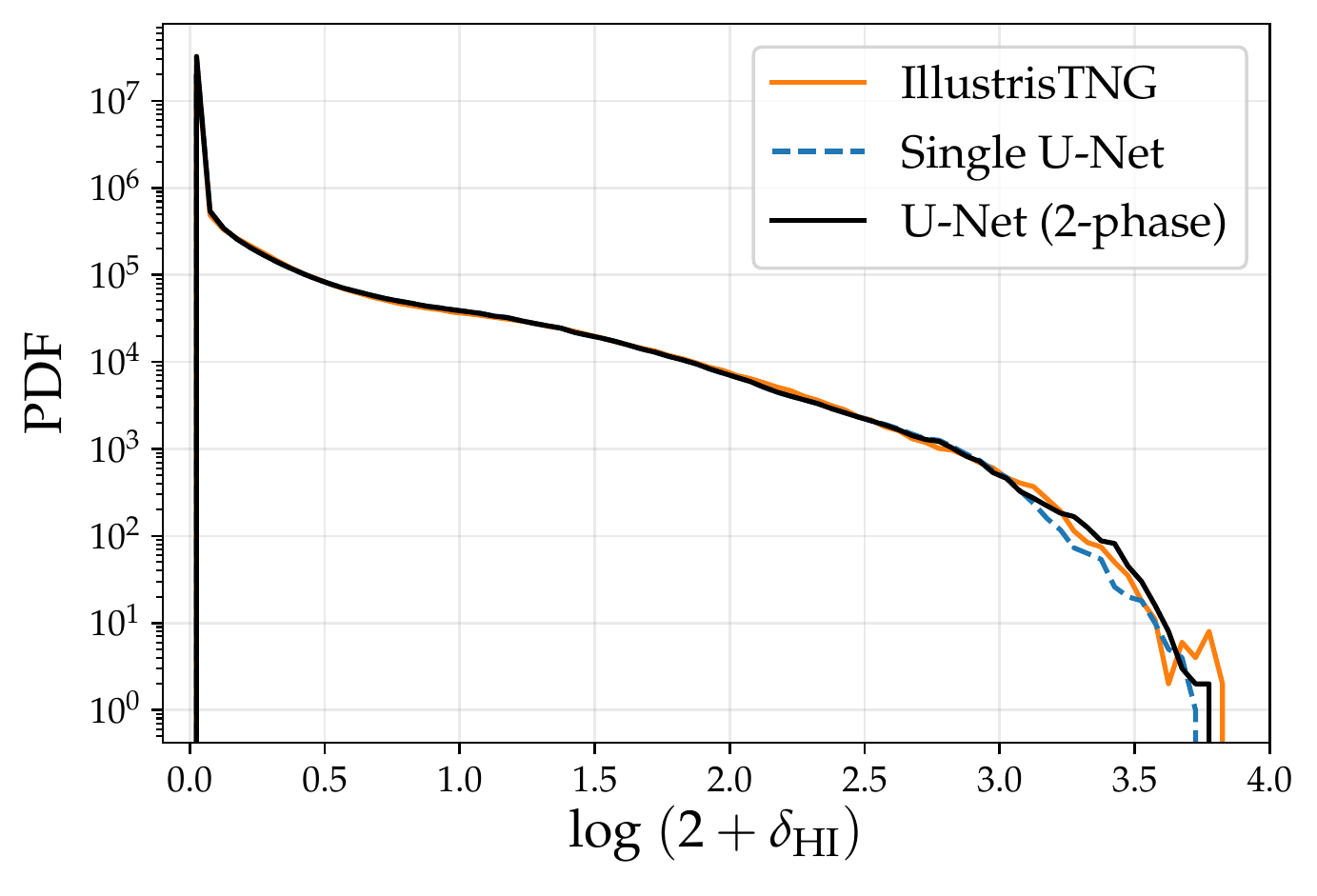}
\includegraphics[scale=0.58,keepaspectratio=true]{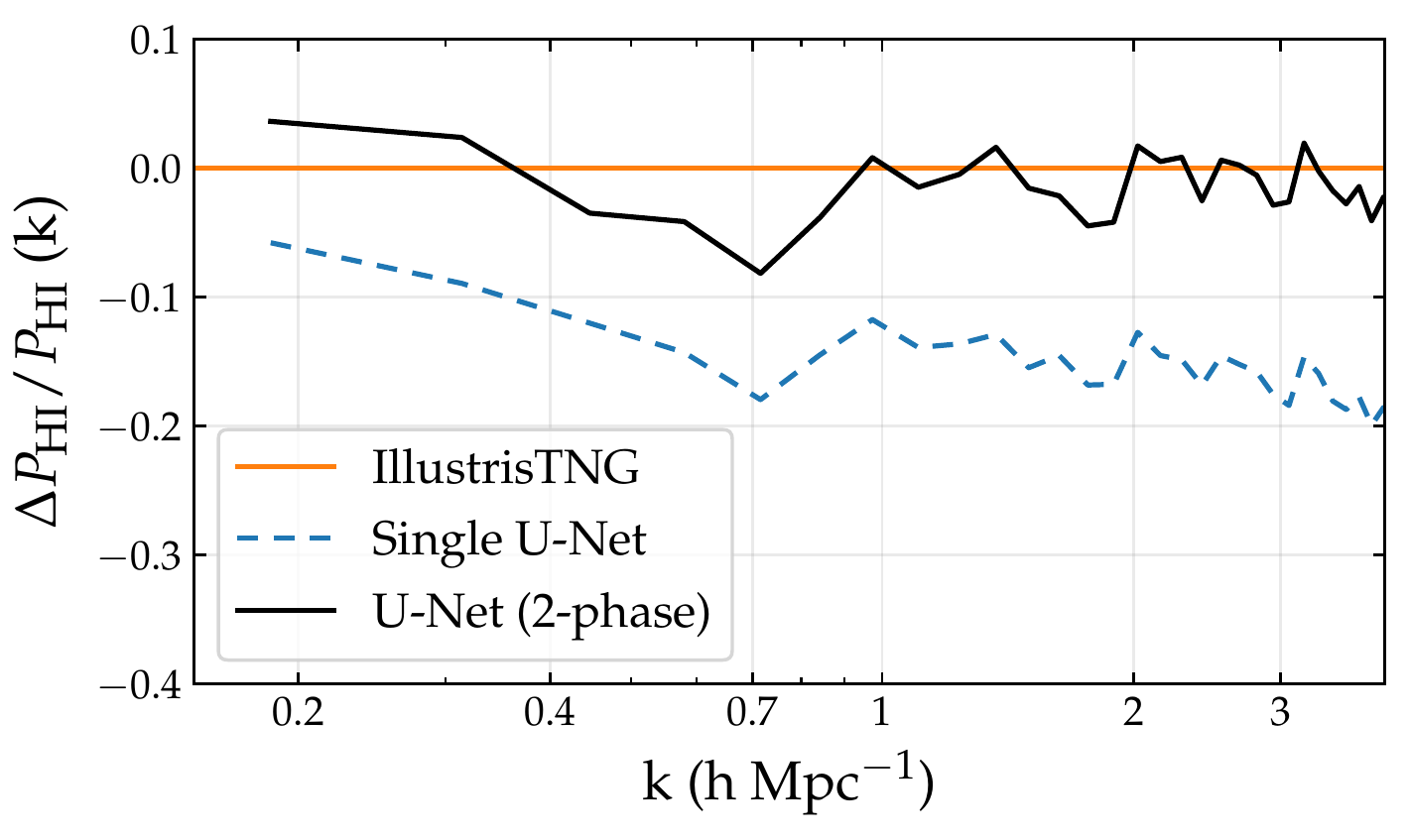}
\caption{Comparison of histogram (left) and the power spectrum (right) on using only a single U-Net and a combination of two U-Nets (2-phase). The combination is designed to outperform a single U-Net for the high-mass \HI voxels.}
    \label{fig:2-phase}
\end{figure*}

The contracting path follows the typical architecture of a convolutional neural network and consists four primary blocks. Each block consists of two successive convolutions with stride 1 and a down-sampling convolution with stride 2, each of the three is followed by batch normalization (BN) and a rectified linear unit (ReLU). 
%The blocks are similar to the network used by \cite{HeLiFen1907}.
For each convolutional layer, we use 3 $\times$ 3 $\times$ 3 filters. As our input DM fields are not periodic, we cannot use periodic padding similar to \cite{HeLiFen1907}; we instead apply a zero padding with size 1. At each down-sampling step we double the number of feature channels and reduce the number of grid-points along each dimension by a factor of 2, and vice versa for the up-sampling step.

We only have one up-sampling step in our architecture. We concatenate the up-sampled map (dimensionality $8^3$) with two maps from the contracting path (denoted with filled boxes): one of the same dimensionality ($8^3$) and an another of a higher dimensionality ($16^3$), but after application of a max-pooling layer. These concatenations help to train the network faster as the gradients are passed through the transverse connections. Concatenation also provides the network with various levels of granularity for the final prediction. We have also tried using a deeper neural network (which can be obtained by using stride 1 instead of stride 2 while moving between layers in figure~\ref{fig:Architecture}), but we did not find an improvement in our results.

\subsection{Two-phase model}
As discussed earlier in Section~\ref{sec:DataSparsity}, we needed to employ a two-phase model to tackle the problem of data sparsity. Let us now discuss in detail how we employ the two neural networks (labelled as FP (SP) for the first (second) phase hereafter). A simple way to think about the functions of the two networks is the following: FP fills the low \HI mass voxels and identifies boundaries of large \HI halos. SP is then used to assign an appropriate \HI mass profile to the large halos.

Both FP and SP have the same U-Net architecture  which was shown in Figure~\ref{fig:Architecture}. If we only use FP in our prediction, we get the results shown in a dashed line in Figure~\ref{fig:2-phase}.
From the PDF, we see that the FP is reproducing the PDF accurately for all \HI voxels except the high-mass ones. We believe this is due to dearth of training data on the high mass \HI voxels and expect the FP to perform better with more data. To complement FP, we choose to employ a SP which is focused on predicting only the high-mass \HI voxels. The training of the SP is therefore a little different from that of FP. 
Equation~(\ref{eq:DataTransform}) was used to generate rescaled fields to train the FP, but for SP, we use different rescaling transformations:
\beq\begin{split}
\tilde{\delta}^{(2)}_\dHI&=0.03\, (1+\delta_\dHI)^{0.5}\\
\tilde{\delta}^{(2)}_\textup{DM}&=0.0035\, (1+\delta_\textup{DM})^{0.25}\, .
\end{split}\label{eq:DataTransform2}\eeq
This is because the high-mass tails of the rescaled DM and \HI distributions are flatter for the transforms in Equation~\ref{eq:DataTransform2} as compared to Equation~\ref{eq:DataTransform}, which makes the SP easier to train in the high-mass regime.
We had used the loss function in Equation~\ref{eq:Loss} for training the FP. The loss function for training the SP is obtained by substituting $\tilde{\delta}\rightarrow\tilde{\delta}^{(2)}$ in Equation~\ref{eq:Loss} and we find the hyper-parameters values $\beta=0.26$ and $\alpha=2.5$ give the best results for training the SP.

Let us now discuss combining the predictions of FP and SP after they are trained separately. For a given DM input field, let us denote their unscaled prediction of a particular voxel of the output \HI field by $\delta^\textup{FP}$ and $\delta^\textup{SP}$. We get our final result for that particular voxel by combining the predictions using weights: $w^\textup{FP}\, \delta^\textup{FP} +w^\textup{SP}\, \delta^\textup{SP}$; the weights are given by

\beq
\begin{split}
w^\textup{FP}&\equiv\frac{1}{1+\exp \big[ (\delta^\textup{SP}-\delta^\textup{threshold})/100\big] }\\
w^\textup{SP}&\equiv1-w^\textup{FP}\, ,
\end{split}
\eeq
and were chosen such that the output from SP is given more importance for high-mass voxels and vice versa. We have adopted $\delta^\textup{threshold}=1500$ for results in this paper but we checked that our results are not sensitive to the threshold in the range $\delta^\textup{threshold} \in$(1000 , 2000).

\subsection{Training details and data augmentation}
\label{apx:augmentation}

For training the network, we used the Adam optimizer \citep{Adam} with a learning rate ranging between $10^{-8}-10^{-4}$.
The network takes nearly 30 epochs to train and we choose to use a  batch-size of 28 as it was compatible with the memory constraints of our GPUs.

We also train the neural network to recognize polyhedral symmetries like translational and rotational invariance \citep{HeLiFen1907}. We do this by generating multiple instances of a given input DM box by applying various symmetry transformations and then training the U-Net on all the generated cases. This process is often referred to as data augmentation because we are generating multiple training simulations given using a single data realization.
We use all transformations corresponding to the symmetry group of a cube which consists of 48 elements (octahedral group of order 24 with an additional factor of 2 to account for inversion across the origin ($\vec{r}\rightarrow-\vec{r}$)). As the cubes containing high \HI densities are rare in our training set, using data augmentation greatly helps for training in the high-density regime. We also use a popular technique for tackling data-sparsity called non-uniform sampling, where we input the cubes containing high \HI density voxels more frequently during training. In our case, we first rank the cubes with respect to the maximum \HI density inside them, and then increase the frequency of occurrence of the top 1\% cubes by 100 fold. However, even after using data augmentation and non-uniform sampling, we still had to use the two-phase neural network approach as the data sparsity problem in our case is quite extreme.

 We trained all of our models using New York University's High Performance Cluster using NVIDIA Tesla P-100 and P-40 GPUs. 
 
 \subsection{Testing edge-effects while combining HI boxes}
 \label{apx:edge_effects}
\begin{figure}
\centering
\includegraphics[scale=0.58,keepaspectratio=true]{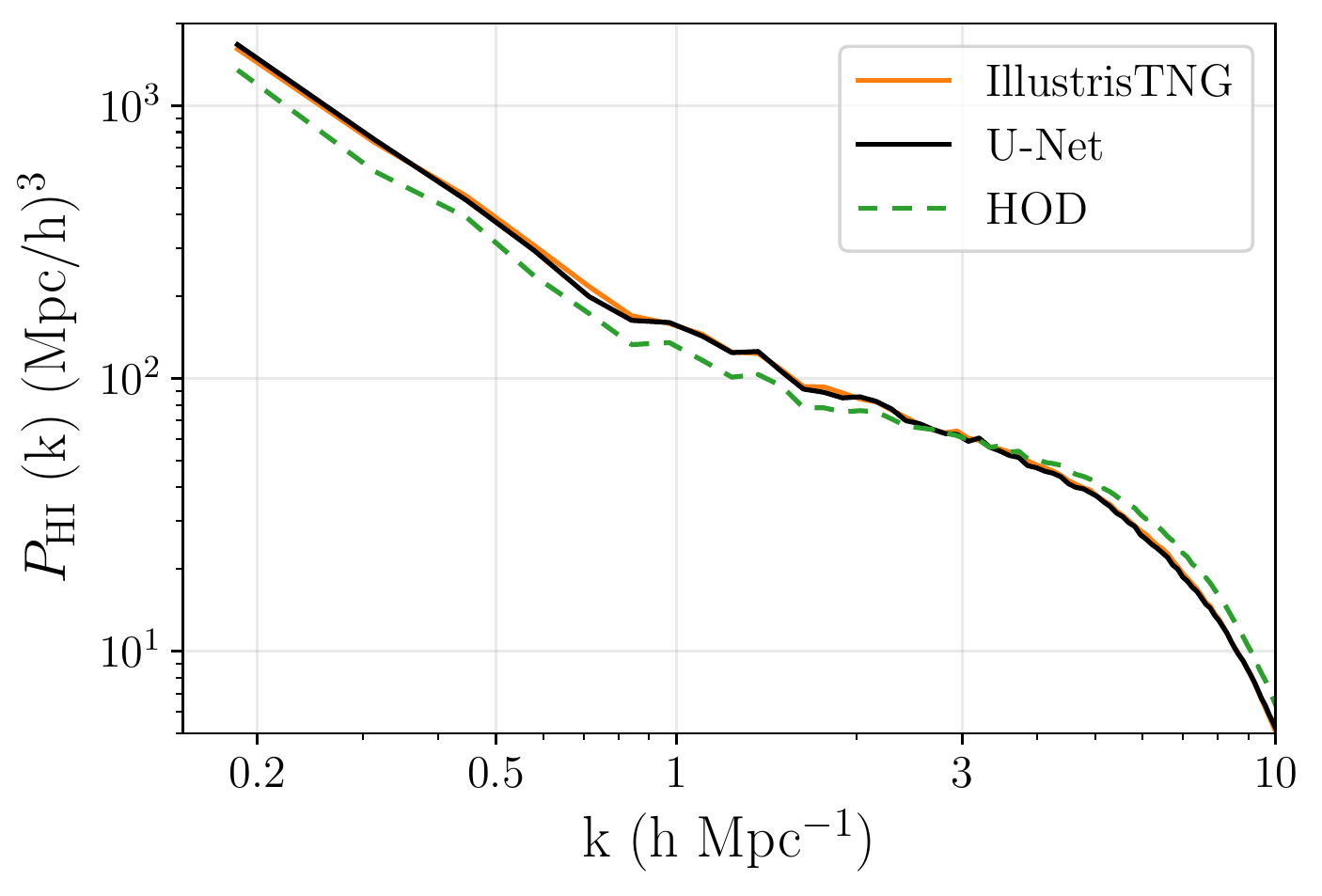}
\caption{Same as the left panel of figure~\ref{fig:Pow} but extended to high-$k$ regions.}
    \label{fig:Pow_k=10}
\end{figure}

\begin{figure}
    \centering
        \includegraphics[scale=0.55,keepaspectratio=true]{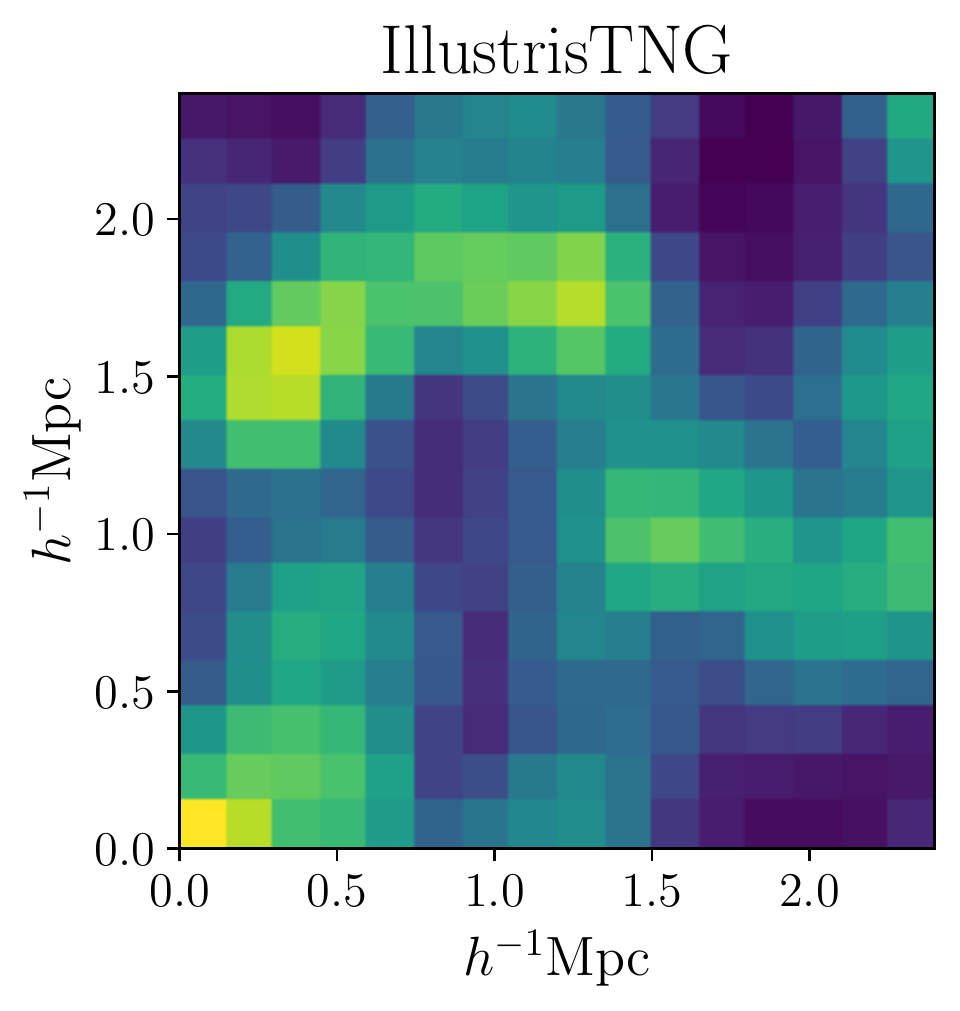}
        \includegraphics[scale=0.55,keepaspectratio=true]{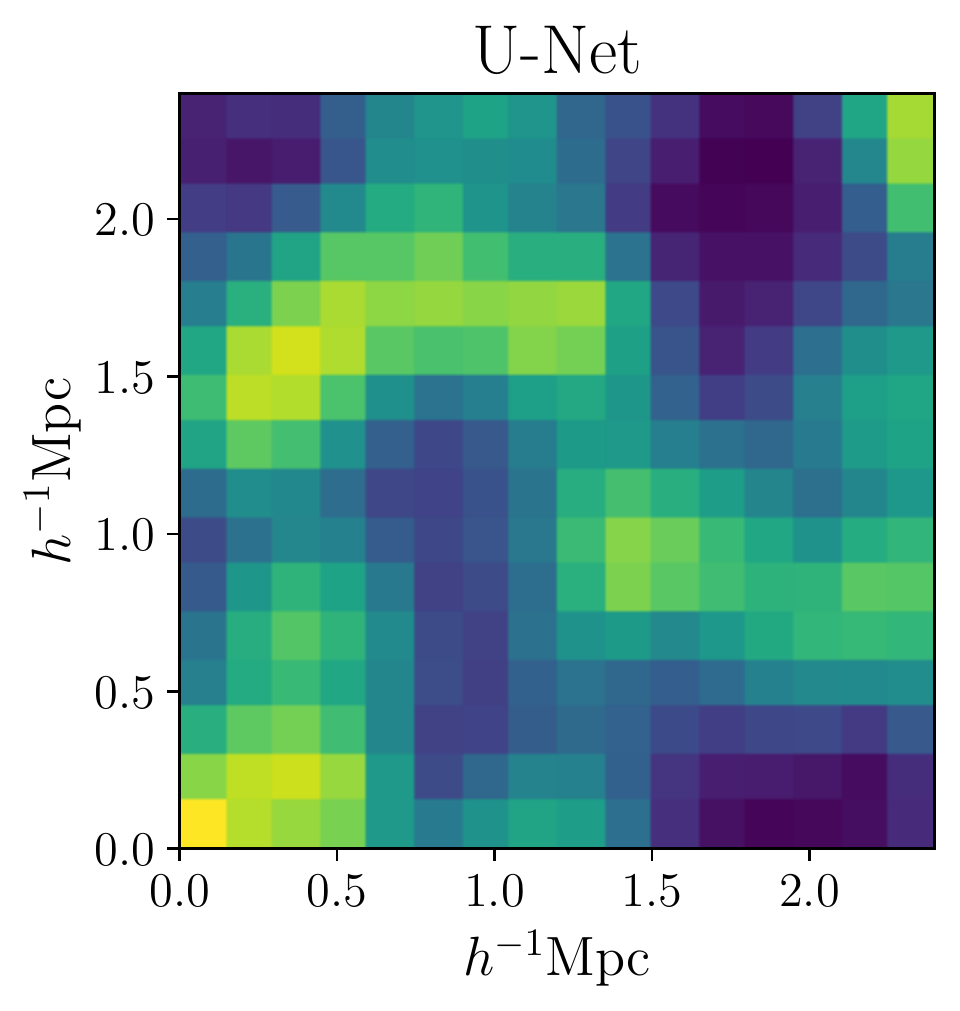}
\caption{Same as the upper panels of figure~\ref{fig:2D} but zooming in on the lower left region of the box in order to verify that there are no spurious edge effects when \HI subcubes of side 1.17$\kMpc$ are concatenated to produce the large U-Net output cube.}
    \label{fig:2D_closeup}
\end{figure}

In section~\ref{sec:results} we showed results from the test set \HI cube. To construct this box, we used the U-Net to generate small $1.17 \Mpc$ side \HI sub-cubes and then stacked them together to produce the test $48 \Mpc$ cube. One possible concern might be that such stacking produces edge effects which will affect summary statistics at scales comparable to the length of the box ($k\sim 5 \kMpc$). We therefore check that there indeed are no artifacts on these scales and show the high-$k$ \HI power spectrum comparison in figure~\ref{fig:Pow_k=10}. 
 Furthermore, we also show a zoom in version of the output of test cube in figure~\ref{fig:2D_closeup}.
A reason for the absence of edge artifacts could be the larger size of DM box used to predict a smaller \HI box (c.f. figure~\ref{fig:Architecture}), so information beyond the location of edges of the \HI box is included in the U-Net prediction.
We also tried using the technique of annealing when concatenating the small \HI subcubes to smoothen their edges, but our results did not change.
%We also checked that generating sub-cubes with overlapping areas and then combining them by averaging over the overlapping areas does not change the summary statistics, indicating that it is a small-scale effect.} 
 
 %We trained for approximately 200 hours .

%Instead of using an analogue of the mean square error for the loss function in Equation~\ref{eq:Loss}, we plan to use the Wasserstein loss in a future work.

% \section{Additional tests}
% We performed the test using the full resolution (35 $\Kpc$) \HI map but the U-Net tends to do poorly.
\end{appendix}
\bibliography{HI}
\end{document}